\documentclass[twocolumn,twocolappendix]{aastex701}
\let\longtable\undefined
\usepackage{newtxtext,newtxmath}
\DeclareRobustCommand{\VAN}[3]{#2}
\let\VANthebibliography\thebibliography
\def\thebibliography{\DeclareRobustCommand{\VAN}[3]{##3}\VANthebibliography}

\usepackage{graphicx}	
\usepackage{amsmath}	
\usepackage{gensymb}
\usepackage{comment}
\usepackage{cleveref}
\usepackage{float}
\usepackage{parskip}
\usepackage{enumitem}

\begin{document}

\title{Zippers and Twisters II: \\
The lopsidedness of satellite systems around local massive galaxies emerges from a shifting cosmic web landscape}

\correspondingauthor{Janvi P. Madhani, Charlotte Welker}

\author[0000-0002-0913-991X]{Janvi P. Madhani$^*$}
\affiliation{Johns Hopkins University, 
3400 N. Charles Street
Baltimore, MD 21218}
\affiliation{NYC College of Technology,
City University of New York,
NY, USA}
\email[show]{jmadhan1@jh.edu}

\author[0000-0001-5576-0144]{Charlotte Welker$^*$}
\affiliation{NYC College of Technology,
City University of New York,
NY, USA}
\affiliation{Graduate Center,
City University of New York,
NY USA}
\email[show]{cwelker@citytech.cuny.edu}

\author{Sneha Nair}
\affiliation{NYC College of Technology,
City University of New York,
NY, USA}
\affiliation{Graduate Center,
City University of New York,
NY USA}
\email[]{snair@gradcenter.cuny.edu}

\author{Katarina Kraljic}
\affiliation{Observatoire Astronomique de Strasbourg, Université de Strasbourg, CNRS, UMR 7550, F-67000 Strasbourg, France}
\email[]{kraljic@unistra.fr}

\author{Christophe Pichon}
\affiliation{Institut d'Astrophsyique de Paris,
UMR 7095, CNRS, UPMC Univ.,
Paris VI 98 bis boulevard Arago,
Paris, France}
\affiliation{Kyung Hee University, Dept. of Astronomy \& Space Science, Yongin-shi, Gyeonggi-do 17104, Republic of Korea}
\email[]{pichon@iap.fr}

\author{Yohan Dubois}
\affiliation{Institut d'Astrophsyique de Paris,
UMR 7095, CNRS, UPMC Univ.,
Paris VI 98 bis boulevard Arago,
Paris, France}
\email[]{dubois@iap.fr}

\begin{abstract}
We investigate the role of the cosmic web in the origin and strength of lopsidedness in the distributions of resolved luminous dwarf satellites around Milky-Way to Centaurus A analogs in the high-resolution hydrodynamic simulation NewHorizon. First, we apply commonly used lopsidedness estimators to compare the degree of lopsidedness of NewHorizon systems to systems presented as anomalies or challenges to $\Lambda$CDM in observations. We find that  $>25\%$ of our sample display ``statistically significant” lopsidedness incompatible with expectations for dark matter sub-haloes and that the NewHorizon sample is generally skewed towards being more lopsided than these naive expectations. This cautions against using dark matter or under-resolved hydrodynamic simulations as a null hypothesis to compare to real systems. Overall, we find broad, hemispheric lopsidedness to be much more prevalent than narrow, focused lopsidedness simply inherited from group infall. We further find that, across all systems, the dividing plane between the richest and emptiest hemispheres is systematically aligned with a contrasted nearby, intermediate-scale cosmic filament. Further analysis reveals that the lopsidedness of the satellites' distributions is in most cases inherited from the lopsidedness of the network of cosmic filaments feeding it and depends on how the system is positioned in this network. For instance, the most populated side contains the highest number of filaments in all but one case (characterized by a unique host filament that has simply shifted over time). In most cases, this anisotropy of the network is also indicative of an ongoing or recent zipper (edge-on filament merger $<2$ Gyr). Twister remnants (highly helicoidal remnants of high orbital momentum mergers of filaments) are also found on the richest hemisphere of our systems, consistent with the fact that high vorticity regions might help prevent satellite mergers, hence better allowing for the pooling of satellites. Converging filamentary structures therefore funnel satellites preferentially onto one side of host halos, naturally generating lopsided configurations without requiring a recent major galaxy merger event. These results suggest that satellite lopsidedness is a common outcome of the evolving nature of intermediate scales of the cosmic web. As such, it is better understood as a natural consequence of filamentary accretion rather than solely as a transient signature of halo mergers. This result also highlights the need for increased hydrodynamic resolution when simulating these regions.

\end{abstract}



\section{Introduction} 
Observations in the nearby Universe reveal that satellite galaxies around massive hosts can exhibit significant departures from isotropy, including a preferential occupation of one side of their host galaxy or preferential clustering within a narrow cone, referred to as lopsidedness \citep{Libeskind_2016,Brainerd_2020}. In several well-studied systems, most notably Andromeda (M31), satellites preferentially occupy a single hemisphere, often oriented towards a nearby massive companion \citep{Libeskind_2016,Kanehisa_2025}. Statistical studies using the Sloan Digital Sky Survey (SDSS), as well as observations from the MATLAS and ELVES surveys, have further established that lopsidedness is not restricted to interacting galaxy pairs but is also present among isolated host galaxies \citep{Brainerd_2020,Heesters_2024}. These studies reveal a statistically significant excess of satellites occupying the same side of their hosts, both in stacked populations and in a subset of individual satellite systems.

Within $\Lambda$CDM, anisotropic accretion along the cosmic web naturally produces departures from isotropy in satellite populations. Subsequent orbital evolution within host halos can weaken these anisotropies, although continued accretion may sustain them \citep{Liu_2024}. The presence of a nearby massive companion can further enhance lopsidedness, with observational and numerical studies finding preferential satellite distributions towards the region between galaxy pairs \citep{Libeskind_2016,Pawlowski_2017,Gong_2019, Guo2025, 2025ApJ...992...41M}. However, the detection of lopsidedness around isolated hosts suggests that interactions between massive companions are not a necessary condition for its emergence, motivating a more detailed investigation of the role of the surrounding cosmic environment.

Recent studies have further examined whether these mechanisms can account for the pronounced asymmetry of the satellite distribution around M31. Of its 37 known satellites considered by \citet{Kanehisa_2025}, all but one lie within an opening angle of $107^\circ$ towards the Milky Way. Comparing this configuration with M31-like systems in cosmological simulations, \citet{Kanehisa_2025} find that fewer than $0.3\%$ of simulated systems exhibit comparably significant asymmetry, increasing to $0.5\%$ when accounting for possible observational incompleteness. While the orientation of M31's lopsidedness is consistent with the preferential alignment towards a massive companion, its strength and spatial coherence are uncommon among the simulated analogues examined in that study.

The issue is not that $\Lambda$CDM forbids anisotropy \citep{Aubert_2004,2014MNRAS.443.1274L}, but that, the degree of lopsidedness is stronger than expected and suggests kinematic coherence that persists longer than simple phase-mixing arguments would allow for. Together, these conditions challenge assumptions about how satellites are accreted, how quickly they relax dynamically, and how baryonic physics reshape subhalo populations. Several explanations have been proposed over the years. Simulation studies \citep{2008MNRAS.385.1365L, Bowden_2014,2024A&A...687A.212J} show that a significant fraction of satellites are accreted in pairs or groups, in the process of group infall, which statistically increases the chance of spatial and kinematic correlations seen today but does not explain Andromeda's lopsidedness or the general ubiquity of lopsidedness. Many studies investigate the role of baryonic physics and feedback in shaping satellite galaxy distributions and influencing lopsidedness in cosmological simulations \citep{2017MNRAS.466.3119A, Pawlowski_2017}. Indeed, such an investigation reveals that baryonic physics might enhance or suppress anisotropies in satellite populations but fall short of isolating baryonic feedback as a primary driver of lopsidedness. 

A major limitation of previous numerical analysis is likely the limited resolution, in particular in the cosmic environment of such systems. Indeed, observationally and in simulations likewise, satellites show significant alignment with cosmic filaments. Filamentary accretion is a known driver of anisotropic spatial distributions of satellites and therefore could manifest as lopsidedness \citep{Tempel_2015, wang2020alignmentsatellitesystemscosmic,Aubert_2004}. But in large cosmological simulations, both tidal and hydrodynamic resolution quickly degrades outside of galaxies, limiting the degree of anisotropy baryons can inherit from their environment. This limitation is all the more dire for baryons as, unlike dark matter, cosmic gas is in dissipative evolution. Hence lack of resolution in gas features can significantly underestimate their strong potential for thinning through shocks and cooling.

The advent of so-called ``cosmological zooms" like NewHorizon \citep{2021A&A...651A.109D}, combining 34 pc resolution in galaxies ensuring a proper multiphase interstellar medium, but also cosmic volumes and  preserved high {\it environmental} resolution  ($<15$ kpc) out to Megaparsecs from galaxies, is now allowing us to explore in more details the smaller scales of the cosmic web, their evolution and their impact on systems of satellites. For instance, in our recent study \citep{Madhani2026Zippers}, we found that planes of satellites around analogs of local massive galaxies emerge naturally in NewHorizon due to constraints imposed by local cosmic filaments of intermediate scales.  In particular, our study highlighted the role of zippers (filament mergers) and thin tidally-stirred gas streams in their midst, not resolved in larger simulations, in the formation of such planes. The slow evolution of such cosmic filaments, the primary channels of satellites into halos, is therefore a promising avenue to explain satellite lopsidedness, if such mechanisms lead to satellites being accreted preferentially from one side. However we also demonstrated that these pronounced anisotropies in the distribution of dwarf satellites (like co-rotating planes) tend to be mediated by similarly stark geometrical anisotropies in the thin ($<15$ kpc) streams of gas that feed these luminous satellites in the first place. This explains why only simulations with high enough hydrodynamic spatial resolution like New Horizon are able to reproduce degrees of anisotropies similar to observations. One can then expect that resolved runs like New Horizon might also be necessary to quantitatively reproduce other characteristic anisotropies of satellite distributions, like their lopsidedness.

In the present study, we therefore first quantify the degree and ubiquity of lopsidedness among local massive galaxies analogs in NewHorizon, compare it to observed systems, then we investigate the correlation between lospidedness and the dynamics of the gas-rich cosmic web.

Our study is structured as follows: in Section~\ref{sec:methods1} we describe the simulation and general numerical methods used in the study, in Section~\ref{sec:methods2} we detail the quantitative and statistical methods used to define and assess lopsidedness and its correlations with the cosmic web. Then in Section~\ref{sec:census} we present a census of lopsidedness of luminous satellites in NewHorizon, highlighting in particular its ubiquity and extreme nature compared to dark-matter based expectations. In Section~\ref{sec:cweb}, we explore the cosmic origin of lopsidedness and in particular its link to the evolution of the cosmic web. Finally, in Section~\ref{sec:conclusion}, we discuss our results.


\section{Numerical Methods}
\label{sec:methods1}

In this section, we describe the technical specifications of the NewHorizon simulation and the numerical methods used to identify galaxies, halos and cosmic filaments.

\subsection{NewHorizon Simulation}
NewHorizon is a high-resolution hydrodynamic cosmological zoom-in simulation run with \texttt{RAMSES} \citep{2002A&A...385..337T}, centered on a 20 Mpc in diameter co-moving sphere of average cosmic density. It is re-simulated from the Horizon-AGN \citep{2014MNRAS.444.1453D} simulation which has kpc-resolution. We summarize its main properties here \citep[see][for details]{2021A&A...651A.109D}. NewHorizon follows $\Lambda$CDM cosmology, with $\Omega_{\rm m}$ = 0.272, $\Omega_{\Lambda}$ = 0.728,  $\Omega_{\rm b}$ = 0.045, and $\sigma_8$ = 0.81,  $H_0$ = 70.4 km s\textsuperscript{-1} Mpc \textsuperscript{-1} and $n_s$ = 0.967 following \cite{2011ApJS..192...18K}. The dark matter and stellar particle masses are $m_{\rm DM}$ = 1.2 $\times 10^{6} \, {\rm M_\odot}$ and $m_{*}= 1.3 \, \times 10^{4} \, {\rm M_\odot}$ respectively, with a quasi-Lagrangian AMR scheme reaching a maximal spatial resolution of $\Delta x \approx$ 35 pc.  NewHorizon includes detailed gas cooling processes reaching below 10\textsuperscript{4} K, a uniform UV background after z\textsubscript{reion} = 10, turbulent star formation, stellar feedback \citep[Type Ia/II supernovae and winds:][]{2015MNRAS.451.2900K}, and bimodal, spin-dependent AGN feedback (quasar and jet modes).

NewHorizon resolves low mass satellite galaxies, dwarfs, down to stellar masses of $M_* \approx 10^{5.8} \mathbf{M_\odot}$ with at least 50 stellar particles, amounting to more than 5200 dwarfs at redshift $z<0.2$. Galaxies with ${M_* > 10^{6.5} \mathbf{M_\odot}}$ are resolved with at least 240 particles, allowing for accurate reconstruction of the dwarfs angular momentum and structure. Notably, the resolution in filaments several Mpc away for more massive galaxies remains within 35 kpc, allowing for detailed reconstruction of gas filaments in the cosmic web across multiple scales.

\subsection{Halo and Galaxy Identification}

\begin{figure}
	\includegraphics[width=\linewidth]{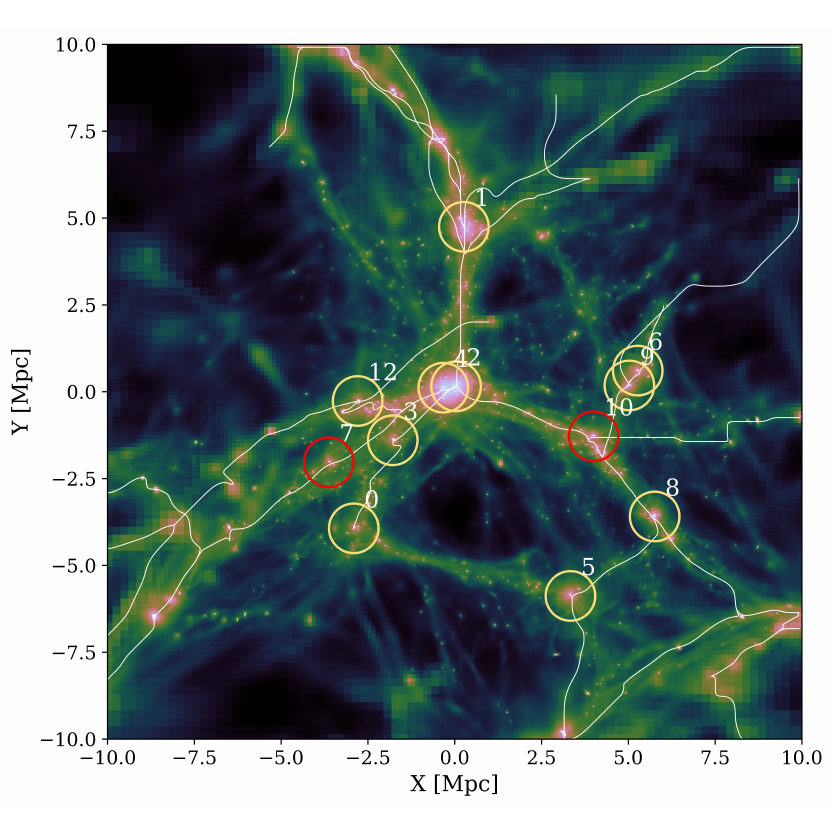}
    \caption{The 12 MW mass systems at $z= 0.17$ overlaid as colored circles on the full projected gas density map of NewHorizon. Systems that are satellite rich ($N_{\rm sat} \geq 14$) are shown with yellow circles and those that are less rich ($N_{\rm sat} < 14$) are shown in red. White lines indicate large-scale cosmic filaments.}
    \label{fig:mainmap}
\end{figure}

We identify halos and galaxies using the density-based structure finder \texttt{HALOMAKER} \citep{2009A&A...506..647T}, a modified 3D friend-of-friend program using the \texttt{AdaptaHOP} algorithm \citep{Aubert_2004} to identify both structures and substructures. Galaxies are identified from stellar particles down to $6.5\,\times10^{5}\, {\rm M_\odot}$ (50 particles), while halos are identified using DM particles down to $7\,\times10^{7}\, {\rm M_\odot}$ (50 particles).

 We identify local massive galaxy analogs (Milky-Way to Cen A mass) as any halo with $ 10^{11.5} {\rm M_\odot} \leq M_{\rm DM} \leq 10^{13} {\rm M_\odot}$ containing a central galaxy with $M_* > 10^{10.5} {\rm M_\odot}$. The higher mass end allows to include analogs of systems like Centaurus A. The same selection criterion were used to study planarity in \cite{Madhani2026Zippers}. In addition, we restrict the analysis to systems with at least 7 identified luminous satellites within the virial radius $R_{\rm vir}$ of the halo, which effectively excludes System 11 from \cite{Madhani2026Zippers} as it hosts only 4 satellites, an insufficient number to properly assess lopsidedness. Note that luminous satellite refers to a clearly identified satellite galaxy ($M_{*}>6.5\,\times10^{5}\, {\rm M_\odot}$). As explained in \cite{Madhani2026Zippers}, this avoids counting spurious, under-resolved stochastic star formation in DM halos as satellites.

For every system, we identify all luminous satellite galaxies that are within 2 $R_{\rm vir}$ of the host DM halo. We calculate the virial radii for our systems as $R_{\rm 200}$ (200 times critical density) and they typically range from 200 to 330 kpc, with only the three most massive systems 1, 2 and 4 ($M_{*}>10^{11.3} {\rm M_{\odot}}, M_{\rm DM}>10^{12.7} \, \rm M_{\odot}$ ) between 350 and 450 kpc.

We look at the level of substructure associated with the central halo, excluding all substructures of the central or another satellite from the analysis to prevent the artificial boosting of any lopsided signal. Most satellites are dwarf galaxies with $10^{6.5} \mathbf{M_{\odot}}< M_{*} <10^9 \mathbf{M_{\odot}}$. We control that the level of contamination from low-resolution particles within 5 $R_{\rm vir}$ is $<5\%$. With this set of parameters, we identify 12 local massive galaxy analogs in NewHorizon at $z<0.2$, shown in Figure \ref{fig:mainmap}. We note that this is one less system than used in \cite{Madhani2026Zippers} as we dropped System 11. However, we choose to preserve the indexing of our previous study for easier comparison, hence the presence of a ``System 12" in the present analysis.

\subsection{Cosmic filament extraction}

We extract the large-scale cosmic filaments by using \texttt{DisPerSE} \citep{2011MNRAS.414..384S} software. \texttt{DisPerSE} is a topology-based structure finder that detects filaments by tracing the ridge lines of a density field, which may be constructed either from discrete tracers such as galaxy catalogs or from a continuous field sampled on a regular grid. Each ridge line links a local density maximum to a filament-type saddle point, to form a “critical pair.” To remove spurious structures, the method applies a persistence threshold that filters out insignificant filaments. Persistence, defined here as a signal-to-noise ratio, measures the density contrast associated with each critical pair. The resulting output is a fully connected filament network represented by contiguous line segments (or equivalently, sampling points) that connect density peaks to filament-type saddle points. 

In the present study, we extract filaments directly from the smoothed gas density field. We apply \texttt{DisPerSE} on a volume of $(20 \, \rm Mpc)^3$ centered on NewHorizon and sample the density field on a $175^3$ pixel box ($\Delta x_0= 114$ kpc). We smooth the density field using a Gaussian filter with a standard deviation of 7 pixels. From here, we pass the field to \texttt{DisPerSE} to extract the largest, high persistence ($5\sigma$) cosmic filaments with a typical length of 2-6 Mpc and $< 1\, \rm Mpc$ in thickness. This network connects nodes with typical MW mass halos ($>10^{12.5} \, \mathbf{M_{\odot}}$) and corresponds to the largest cosmic filaments identified in \cite{Madhani2026Zippers}. Note however that, within the broader scope of cosmic web analysis, these filaments are better described as {\it intermediate-scale} cosmic filaments, typically shorter and thinner than filaments connecting  groups and clusters $>10^{13.5}\rm M_{\odot}$ often identified in large surveys \citep{Tempel_2015,Welker20,Kraljic21,Barsanti2022}.

\section{Lopsidedness: Quantitative methods}
\label{sec:methods2}

This section summarizes the methods we use to characterize and define the lopsidedness of simulated satellite galaxy distributions in NewHorizon. We also introduce the statistical methods by which we assess whether the lopsidedness of these luminous systems significantly deviates from dark matter expectations.

\subsection{Plane of maximum asymmetry}
\label{sec:plane_max_asym}

\begin{figure}
	\includegraphics[width=1.1\columnwidth]{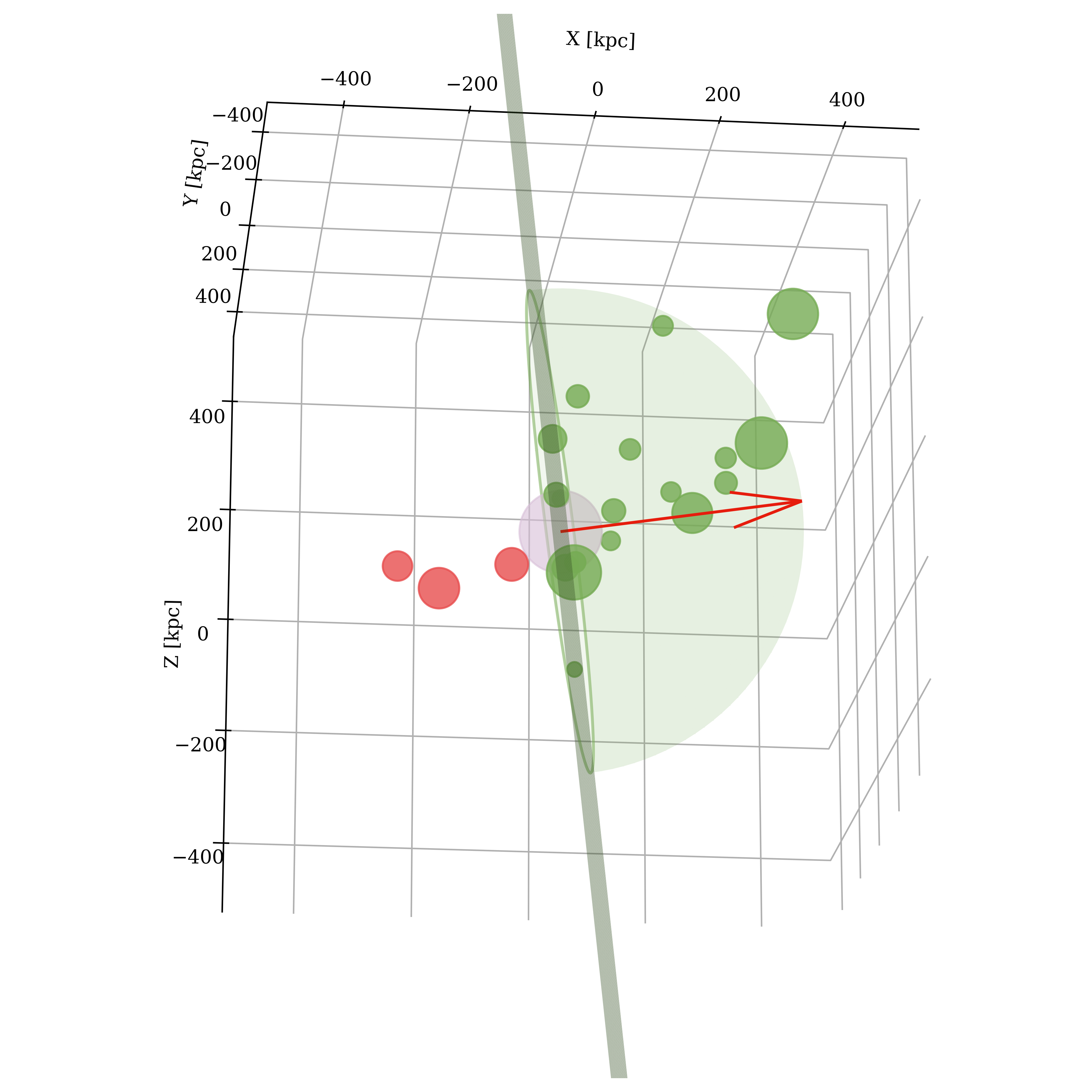}
    \caption{Distribution of the 21 satellite galaxies in System 6, with the central galaxy shown in light purple. The plane of maximum asymmetry (dark green, with its normal vector indicated by the red arrow) divides the satellite distribution into two hemispheres with the greatest difference in satellite counts. The translucent green hemisphere highlights the more populated side of the dividing plane. Satellites are color-coded green and red according to their hemisphere, illustrating the pronounced spatial asymmetry. }
    \label{fig:planemaxasymm}
\end{figure}

\begin{deluxetable*}{ll}

\tablecaption{Description of metrics used to characterize satellite lopsidedness. \label{tab:lop_metrics}}

\tablehead{
\colhead{Metric} & \colhead{Description}
}

\startdata
$N_{\rm lop}$ & Number of satellites in the most populated hemisphere separated by the plane of maximum asymmetry. \\
$R_{\rm lop}$ & Ratio of satellites in the most populated hemisphere to the  total number. \\
$\vec{n}_{\rm lop}$ & Normal vector defining the plane of maximum asymmetry. \\
$\theta_{\rm lop}$ & Mean pairwise satellite angle, $\langle{\theta}_{ij}\rangle$. \\
$N_\theta$ & Maximum number of satellites enclosed within a cone at a given opening angle. \\
$\theta_{\rm open}$ & Opening angle defining a cone. \\
$\vec{r}_{\rm cone}$ & Vector orientation of the cone axis. \\
$\theta_{\rm cosmic}$ & Angle between the plane of maximum asymmetry and the maximally aligned nearby cosmic filament. \\
$\theta_{\rm plane}$ & Angle between the filament-plane normal and the normal of the plane of maximum asymmetry. \\
$\theta_{\rm NN,plane}$ &  Angle between $\vec{n}_{\rm lop}$ and the nearest-neighbor direction. \\
$\theta_{\rm NN,fil}$ &  Angle between maximally aligned cosmic filament and the nearest-neighbor direction. \\
\enddata

\end{deluxetable*}

We characterize the hemispheric/planar lopsidedness of satellites by identifying the plane of maximum asymmetry. This is the plane which maximizes the excess of satellites on one hemisphere compared to the other. In other words, the plane of maximum asymmetry is defined as the plane, centered on the host galaxy, that separates the total satellite distribution into two hemispheres and maximizes the number of satellites in one hemisphere. In order to identify this plane, we draw 10,000 randomly generated plane normal vectors uniformly sampled on a sphere centered on the host. We count the number of satellites that lie in each hemisphere separated by the plane and retain the normal vector that maximizes the ratio between the satellite population in one hemisphere with the total number of satellites in the system. We refer to the number of satellites that occupy the most populated hemisphere as \textbf{$N_{\rm lop}$} and the plane defined by the normal vector, $\overrightarrow{n}_{\rm lop}$, that achieves this maximum disparity as the \textbf{plane of maximum asymmetry}. An example of this plane is shown in Figure~\ref{fig:planemaxasymm}. These definitions and properties are summarized in Table \ref{tab:lop_metrics}.

We have tested convergence of $N_{\rm lop}$ with respect to the sampled orientations, finding that $N_{\rm lop}$ (and hence $R_{\rm lop}$) is stable and converged for our adopted sample with 10,000 trial vectors. Note however that, in principle, the orientation of the plane of maximum asymmetry itself is not necessarily uniquely constrained for a finite satellite population. For systems with low numbers of satellites, variations in plane orientation may preserve the same hemispheric satellite count. However, we checked that:
\begin{itemize}[leftmargin=*]
    \item there is no discrete/global degeneracy in the identification of the plane of maximal asymmetry for any system in our sample. The only type of ``degeneracy" we observe corresponds to a family of planes locally spanning a narrow angular continuous range of orientations that all identify the same hemispheric splitting.
    \item Even in this case, we checked that most systems are non-degenerate to $<5^\circ$ precision and all but Systems 7 and 10 are non-degenerate to well under $10^\circ$ precision. We randomly pick a representative among these planes for each system.
    \item Only System 7 and 10, with less than 10 satellites each, show planes of maximal asymmetry that can span a broader range of angles, although it remains $<15-20^\circ$. However, we note that taking the average orientation rather than a random one in these cases actually increases significantly the degree of alignment of the plane with the cosmic web, as illustrated in Appendix~\ref{appendix:degeneracy}. For consistency, we nevertheless present results in the main text for randomly picked representatives.
\end{itemize} 

 Thus, the choice of maximizing orientation introduces only limited uncertainty below our angular resolution for sufficiently satellite rich systems, and does not materially affect our characterization of satellite lopsidedness.

\subsection{Pairwise Angle}

\begin{figure}
	\includegraphics[width=1\columnwidth]{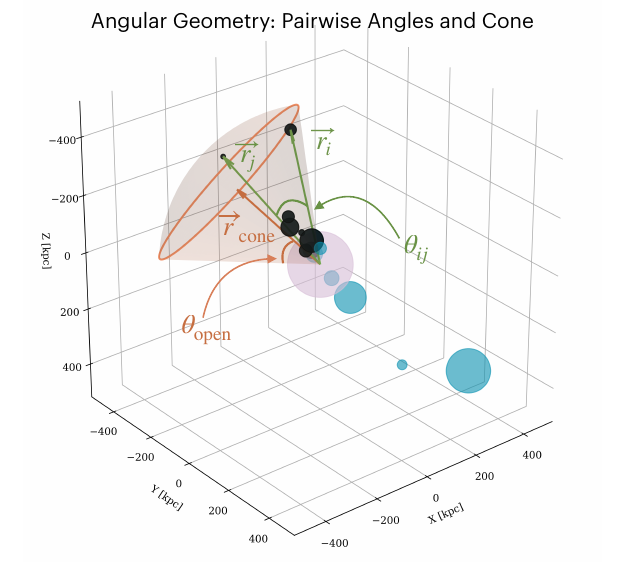}
    \caption{Angular geometry of the satellite distribution in System 12, with the central galaxy shown in light purple. The translucent orange cone, with opening angle $\theta_{\rm open}=40^\circ$, is oriented along $\vec{r}_{\rm cone}$ to maximize the number of enclosed satellites, $N_\theta$, over 1000 isotropically distributed test directions. In this configuration, $N_\theta=8$ of 14 satellites, with those inside and outside the cone shown in black and blue, respectively. The green vectors $\vec{r}_i$ and $\vec{r}_j$ connect the central galaxy to two randomly selected satellites, defining their pairwise angular separation, $\theta_{ij}=29^\circ$.}
    \label{fig:pairwise_and_opening_angle}
\end{figure}

Next, we probe how asymmetric/clustered the spatial distribution of satellites is without requiring any assumption on hemi-sphericity of lopsidedness. In an isotropic satellite system, the angle between any two satellites (as seen from the host) is uniformly distributed in cosine. That means one can expect just as many wide-separation pairs as close-aligned pairs. If the system is lopsided, satellites preferentially occupy one side of the halo, so many pairs will lie in similar directions, producing an excess of small pairwise angles (and a deficit near $180^\circ$).

The pairwise satellite angle, $\theta_{\rm lop}$ is a geometric statistic commonly used to quantify lopsidedness and anisotropy in satellite systems. Consider two satellites, $i$ and $j$ of the same host galaxy. Let $\hat{r}_i$ and $\hat{r}_j$ be the unit separation vectors from the host to each satellite, as illustrated in Fig.~\ref{fig:pairwise_and_opening_angle}. Then, the pairwise satellite angle is: 

$$\theta_{ij} \equiv \rm arccos(\hat{r}_i \cdot \hat{r}_j)$$ 

\begin{itemize}
    \item $\theta_{ij} = 0^\circ$: satellites are in the same direction 
     \item $\theta_{ij} = 180^\circ$: satellites  lie on opposite sides of the host
\end{itemize}

For an isotropic satellite distribution, we expect the probability distribution of $\rm cos (\theta_{ij})$ to be uniform in $[-1,1]$. If satellites are lopsided, many satellites occupy the same hemisphere or the same cones. Therefore pairs are more likely to have $\theta_{ij} < 90 ^\circ$, creating an excess in the probability distribution for specific angle ranges and a shift in the ensemble mean. Crucially, this signal appears without choosing a preferred direction. We define $\theta_{\rm lop}$ as the mean pairwise satellite angle, $\langle   \theta_{ij} \rangle$. In a perfect isotropic distribution $\theta_{\rm lop}=90 ^\circ$.

\subsection{Cones - Solid Angles}
\label{sec:cone_solidangles}

Using the hemispheric method and mean pairwise satellite angle provides a good overall indicator of the lopsidedness of satellite populations, but fails to identify the more granular angular regions to which satellites can be confined, for example, in clusters. To more robustly quantify the level of asymmetry in satellite distributions, we use the cone method as defined by \cite{Kanehisa_2025}. For each system in NewHorizon, we generate 359 cones defined by an opening angle, $\theta_{\rm open}$, that spans the interval of $[0,180]^\circ$, in $0.5^\circ$ increments. Each cone direction is isotropically rotated 1000 times around the center of the system, defined by the location of the central galaxy. This gives a total of 359 $\times$ 1000 cones with systematically increasing opening angles to calculate asymmetry. For each cone, we count the number of satellites captured within the volume of the cone. For a given opening angle $\theta_{\rm open}$, the cone of maximum asymmetry (defined by its direction $\overrightarrow{r}_{\rm cone}$) is the cone that captures the maximum number of satellites, $N_{\theta}$. We show one such cone in Figure \ref{fig:pairwise_and_opening_angle} to help visualize the method with which $N_{\theta}$ is measured. 

We summarize the various metrics we calculate in Table \ref{tab:lop_metrics}.

\subsection{Statistical significance of lopsidedness}

For each system of satellites, we assess whether the lopsidedness of the satellite distribution represents a significant departure from naive (non-baryonic) $\rm \Lambda CDM$ assembly models, typically used in observations to assess the significance of lopsidedness. We test the plane of maximum asymmetry $\overrightarrow{n}_{\rm lop}$, mean pairwise angle $\theta_{\rm lop}$, as well as the the cone of maximum satellites, against a naive, commonly accepted model for the distribution of satellites: the halo-tracing assumption. We mimic observational studies that typically test the lopsided distribution of satellites of a given system against a null hypothesis extracted from dark matter simulations (or at times from under-resolved hydrodynamic simulations effectively equivalent to DM runs). We test both each single system's lopsidedness significance against the expectation derived from its own DM halo, and the significance of lopsidedness at the population level through the entire NewHorizon sample. The generation of corresponding null-hypothesis is described hereafter.

\subsubsection{Per-Halo Analysis}

For each halo, we compare the asymmetry ratio $R_{\rm lop}=N_{\rm lop}/N_{\rm sat}$, $\theta_{\rm lop}$, and $N_{\theta}$ for the NewHorizon systems to their distributions across a synthetically generated null-hypothesis. 

We generate the null hypothesis distribution by resampling each system's satellite population 5,000 times, preserving the total number of satellites $N_{\rm sat}$, their radial distribution, and the underlying ellipsoidal shape of the host halo. The host halo is approximated as an ellipsoid with principal axes $c<b<a$, determined from the inertia tensor of DM particles within $R_{\rm vir}$. To avoid bias from sub-haloes associated with luminous satellites, DM particles within $R_{\rm vir}^{\rm sub}$ of each sub-halo are removed and replaced by particles sampled from the host's average radial density at the same radius. Satellite directions are then drawn uniformly from the surface of an ellipsoid with axes $a$, $b$, and $c$, normalized to a circularized radius of $R_{\rm circ}=(abc)^{1/3}=1$. Finally, the circularized radius of each satellite $R_{\rm circ}$ is also drawn so as to reproduce the observed radial distribution of satellites in each system.

We conduct the analysis two ways. First, we compare each system's lopsidedness parameters to its null-hypothesis. Then we restrict the analysis to systems with 37 or more satellites, which we repeatedly sub-sample to $N_{\rm sat} = 37$. More specifically, each system is resampled 1000 times and for each iteration we calculate $N_{\rm lop}$, $\theta_{\rm lop}$ and $N_{\theta}$.  The sub-sampling is also applied to the null-hypothesis of each system. This makes our systems more directly comparable to observed system M31 and removes the dependence of lopsidedness statistics on satellite number, ensuring that differences between systems reflect variations in the underlying spatial distribution rather than sample size.

Synthetic systems of the null-hypothesis are also used to assess the significance of cone lopsidedness. For this, we simply measure $N_{\theta}$ (the maximal cone occupation) as a function of cone opening angle for each each synthetic mock. This allows us to assess if and in which angular ranges NewHorizon systems cone anisotropies significantly deviate from underlying DM expectations (see Fig.~\ref{fig:cone_bg} for examples).

\subsubsection{Population Analysis}
\label{sec:pop_analysis}

To conduct an analysis at the population level for MW-mass galaxies and above, assuming that NewHorizon is representative in this mass range, we also generate a composite null distribution of lopsidedness parameters $R_{\rm lop}$, $\theta_{\rm lop}$ and $N_{\theta}$ for the full NewHorizon sample. To do so, we create 12 x 5000 synthetic distributions of satellites, under the already discussed assumption that satellites trace the shape of their halo. Note that we assume that each system's structure (halo shape, number of satellites, and radial distribution) is equally likely to occur (hence that NewHorizon is representative).  

\subsubsection{Cone statistic}
\label{sec:cone_stat}

To quantify the global relevance and significance of cones of high lopsidedness across the sample, we also compress the full angular distributions of $N_{\theta}$ into a single scalar statistic, $Z(\theta)$, defined as the normalized difference between $N_\theta$ and its null expectation value. In other words, we standardize each system relative to its own null.

For each NewHorizon system, we therefore compute: 
$$ Z(\theta) = \frac{N_{\theta} - \langle N_{\theta}^{\rm mock}\rangle}{\sigma(N_{\theta}^{\rm mock})}$$
for all opening angles $\theta$ between 0 and $180^\circ$ where:
\begin{itemize}
    \item $N_{\theta}$ = the observed maximum number of satellites enclosed in a cone of  opening angle, $\theta$
    \item $\langle N_{\theta}^{\rm mock}\rangle =$ mean $N_\theta$ over 5000 mocks from the halo-tracing hypothesis (null)
    \item $\sigma(N_{\theta}^{\rm mock}) = $ standard deviation of the 5000 mocks
\end{itemize}

This construction captures the extremeness of the satellite count relative to the halo-tracing expectation of each system. In a perfectly uniform distribution $Z(\theta)=0$, whereas a lopsided cone producing a significant excess would yield $Z(\theta)>2$ across a range of angles. $Z(\theta)$ therefore assesses lopsidedness of all types in one function. This makes it a robust, model-independent diagnostic that is insensitive to the choice of preferred axis or plane. Comparing the distribution of $Z(\theta)$ values across all 12 systems in NewHorizon therefore provides a straightforward way to assess all types of lopsidedness at once.

\subsection{Environmental analysis}

\subsubsection{Alignment with the cosmic web}

To characterize the local filamentary environment of each system, we first identify the node in the filament network nearest to the central galaxy of each system as the main node. If the main node is found within $R_{\rm vir}$ of the central (hence corresponds to its halo), we define the total number of connected filaments as the  connectivity of the system. If no node is found within $R_{\rm vir}$ of the central, we check that one and only one filament crosses $R_{\rm vir}$ of the central and that it is connected to the main node. In this case, the connectivity is set to one. Note that while 4 of NewHorizon systems are in this configuration, even in these cases the main node is found within 3 $R_{\rm vir}$ of the central hence in the tidal/accretion zone of the system.

For each filament connected to the main node, we compute its average orientation vector, $\vec{f}$, from the directions of its constituent spine segments, providing a representative measure of the filament's overall orientation rather than its local orientation at the position of the central galaxy. We then calculate the angle between $\vec{f}$ and the normal vector of the plane of maximum asymmetry, $\vec{n}_{\rm lop}$, for each connected filament. For each system independently, we identify the filament whose average orientation is most closely aligned with its plane of maximum lopsidedness and define $\theta_{\rm cosmic}$ as the minimum angle between the plane and the this filament, as illustrated in green in Fig.~\ref{fig:theta_nn_fil}. This approach allows us to examine whether the orientation of the satellite lopsidedness is preferentially aligned with one of the larger-scale filaments connected to the host's local cosmic-web environment.

\subsubsection{Nearest Massive Neighbor}

We also investigate the role of the nearest massive neighbor (NN), an important environmental diagnostic of any tidal interactions driving satellite anisotropy. A classic example is the Milky Way - M31 pair whose satellite systems appear to point roughly towards one another \citep{Kanehisa_2025}. 

For each of our systems, we identify the nearest neighboring galaxy (NN) that has a mass $\geq 1/3 \,M_{*}$ of the central (corresponding to a pair merger ratio of $25\%$, hence characterizing a major encounter) and we identify the separation vector $\vec{r}_{NN}$, that points from the central to this nearest massive neighbor. This allows us to compute for each system: $d_{NN}$, the distance to the NN, and $\theta_{\rm NN,plane}$, the angle between the plane of maximal asymmetry and $\vec{r}_{NN}$.

\begin{figure}
    \centering
    \includegraphics[width=\columnwidth]{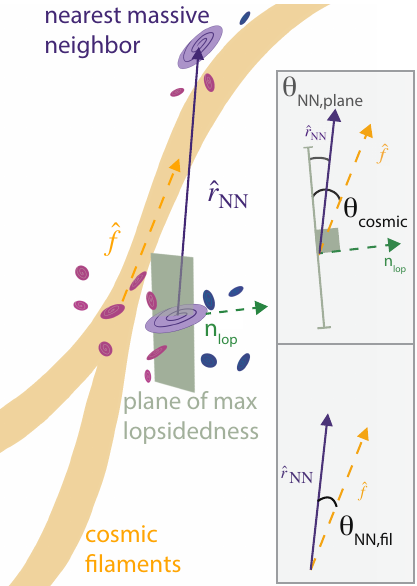} \caption{Illustration of the main angles $\theta_{cosmic}, $ $\theta_{\rm NN, fil}$ and $\theta_{\rm NN, plame} $ used in our environmental analysis. Filaments merging edge-on -- a zipper -- are shown in orange. An example massive galaxy with a lopsided distribution of satellites (pink / blue sides) is shown in purple, its plane of maximum asymmetry in green.}
    \label{fig:theta_nn_fil}
\end{figure}

It should be noted that massive neighbor galaxies often reside within the same cosmic web structures as our target systems (since satellites are accreted preferentially along filaments). Hence, if the nearest massive neighbor lies along the dominant filament, satellites may preferentially enter the host from that direction, naturally creating a lopsided distribution in that direction.
To test the correlations between the direction of the closest filament and the direction of the most massive neighbor, we also calculate the angle $\theta_{\rm NN, fil}$, between the axis of the cosmic filament that is most aligned to the plane of lopsidedness, $\vec{f}$, and the separation vector that points from the galaxy to its nearest massive neighbor, $\vec{r}_{\rm NN}$ (see Fig.~\ref{fig:theta_nn_fil} for a depiction).

In other terms, we calculate: 
$$ \theta_{\rm NN,fil} = \cos^{-1} |\vec{r}_{\rm NN} \cdot \vec{f}|\,\,; \,\,\theta_{\rm NN,plane} = 90^\circ-\cos^{-1} |\vec{r}_{\rm NN} \cdot \vec{n}_{\rm lop}|\,$$

Note that this focus on nearby massive neighbors is chosen to assess the relevance of suggested mechanisms potentially driving lopsidedness in New Horizon, in particular strong torquing from a neighbor in close interaction (future major merger) \citep{Libeskind_2016, Kanehisa_2025}. We are not generally assessing whether satellites are in filaments, which was the topic of Paper I \citep{Madhani2026Zippers}.

\section{Census of Lopsidedness in NewHorizon}
\label{sec:census}

\subsection{Census in NewHorizon}

Let us first quantify the prevalence and strength of lopsided dwarf satellite configurations around our NewHorizon sample of local massive galaxies analogs. To perform this analysis, we use metrics commonly tested in observations, the definitions of which are described in Section \ref{sec:methods2}. In particular, we identify the plane of maximum asymmetry for each of our 12 systems, then calculate the corresponding asymmetry ratio, $R_{\rm lop} = N_{\rm lop} / N_{\rm sat} $, with $N_{\rm lop}$ being the maximal number of satellites found on one side of the plane of maximum asymmetry, as well as $\langle \theta_{ij}\rangle=\theta_{\rm lop}$, their average satellite pairwise angle. 

Note that unlike statistics based on angular correlations, $R_{\rm lop}$ directly measures the degree to which satellites preferentially occupy one side of the host galaxy. This makes it particularly well suited for identifying hemispheric lopsidedness. Because the statistic is normalized by $N_{\rm sat}$, it also allows for easy comparison across hosts with varying numbers of satellites, although one has to account for potential sampling bias when comparing frequencies of given values across systems. On the other hand, the average pairwise satellite angle provides a complementary probe of anisotropy by measuring the angular clustering of satellites independent of a predefined axis and is sensitive to lopsided and filamentary accretion through local clustering. In particular, a hemispheric excess (in $R_{\rm lop}$) can exist without producing a dramatic shift in the pairwise-angle distribution, particularly if satellites are broadly distributed within the favored hemisphere rather than tightly clustered.

Table~\ref{table:lop_table} summarizes these values for all systems, along with their number of satellites $N_{\rm sat}$. It is clear from these numbers that lopsidedness is ubiquitous within this range of mass and richness, with most systems boasting lopsidedness ratios $R_{\rm lop}>75\%$ and half have $R_{\rm lop}>80\%$. Average pairwise angles drop as low as $80.3^\circ$ and several more systems show $\theta_{\rm lop}\leq 87^\circ$. For comparison, Andromeda's satellite population, among the most striking observational examples of lopsidedness, has $\theta_{\rm lop}=83.5^\circ$ and $R_{\rm lop}=80\%$, values comparable to those found in several NewHorizon systems.

\begin{table}
\begin{tabular}{ccccc}
\hline \hline
System & $N_{\rm sats}$ & $N_{\rm lop}$ & $R_{\rm lop}$ & $\theta_{\rm lop} [^\circ]$\\
\hline
0 & 16 & 12 & 0.75 & 91.79  \\
1 & 60 & 42 & 0.70 & 88.24 \\
2 & 84 & 56 & 0.67 & 89.73 \\
3 & 17 & 14 & 0.82 & 90.98 \\
4 & 104 & 78 & 0.75 & 86.59 \\
5 & 26 & 21 & 0.81 & 88.03 \\
6 & 21 & 18 & 0.86 & 87.01 \\
7 & 7 & 6 & 0.86 & 95.88 \\
8 & 32 & 23 & 0.72 & 89.45 \\
9 & 17 & 15 & 0.88 & 90.04 \\
10 & 7 & 7 & 1.00 & 80.29 \\
12 & 14 & 12 & 0.86 & 87.09\\
\hline
\end{tabular}
\caption{Main properties of each system of satellites. This includes: $N_{\rm sat}$ total number of satellites, $N_{\rm lop}$ number of satellites populating a favored hemisphere, the lopsidedness ratio $R_{\rm lop} = N_{\rm lop}/N_{\rm sat}$ ,the average pairwise angle defining the angular clustering of satellite distributions, and $\langle \theta_{ij} \rangle = \theta_{\rm lop}$.} 
\label{table:lop_table}
\end{table}

The relative clustering of systems around $\theta_{\rm lop}=90^\circ$ indicates that most systems do not typically exhibit unusually small pairwise angular separations, yet they possess a relatively strong dipole moment. In other words, satellites are not enclosed in very compact angular wedges, but they are distributed over a broader range of angles towards a net preferred direction. However, two questions arise:
\begin{itemize}[leftmargin=*]
    \item The first one is whether this feature is really an anomaly requiring specific formation mechanisms, potentially of baryonic origin, or whether randomly populating realistic halos at the richness values indicated in Table.~\ref{table:lop_table} naturally produces these levels of lopsidedness.
    \item The second, related one, is whether the extreme lopsidedness of Andromeda can be explained by a $\Lambda CDM$ hydrodynamic simulation and/or whether it may indicate a tension between observed systems and the $\Lambda CDM$ model, as suggested in \cite{Kanehisa_2025} after comparison with simulated DM halos. This is explored in Appendix~\ref{appendix:m31}.
\end{itemize} 

\subsection{Discrepancy with the halo-tracing model}
 
Let's focus on the first question. To characterize how strongly satellite systems depart from the halo-tracing hypothesis, the hypothesis that their distribution merely trace the shape and profile of their host halo, we compare the observed distribution of $R_{\rm lop}$ and $\theta_{\rm lop}$ values in NewHorizon to those of a synthetic distribution obtained by generating 5000 randomized realizations per system. For a given system, each randomized realization is obtained by randomly resampling its DM halo (defined by its inertia tensor) while maintaining $N_{\rm sat}$ and the radial distribution of satellites, then calculating $R_{\rm lop}$ and $\theta_{\rm lop}$.  We remove Systems 7 and 10 from the analysis as they have fewer than 10 satellites. Indeed, due to stochastic sampling effects, low satellite counts can artificially enhance measured lopsidedness. Note that all remaining systems have $N_{\rm sat}\geq 14$.

\begin{figure}
\centering
	\includegraphics[width=1\columnwidth]{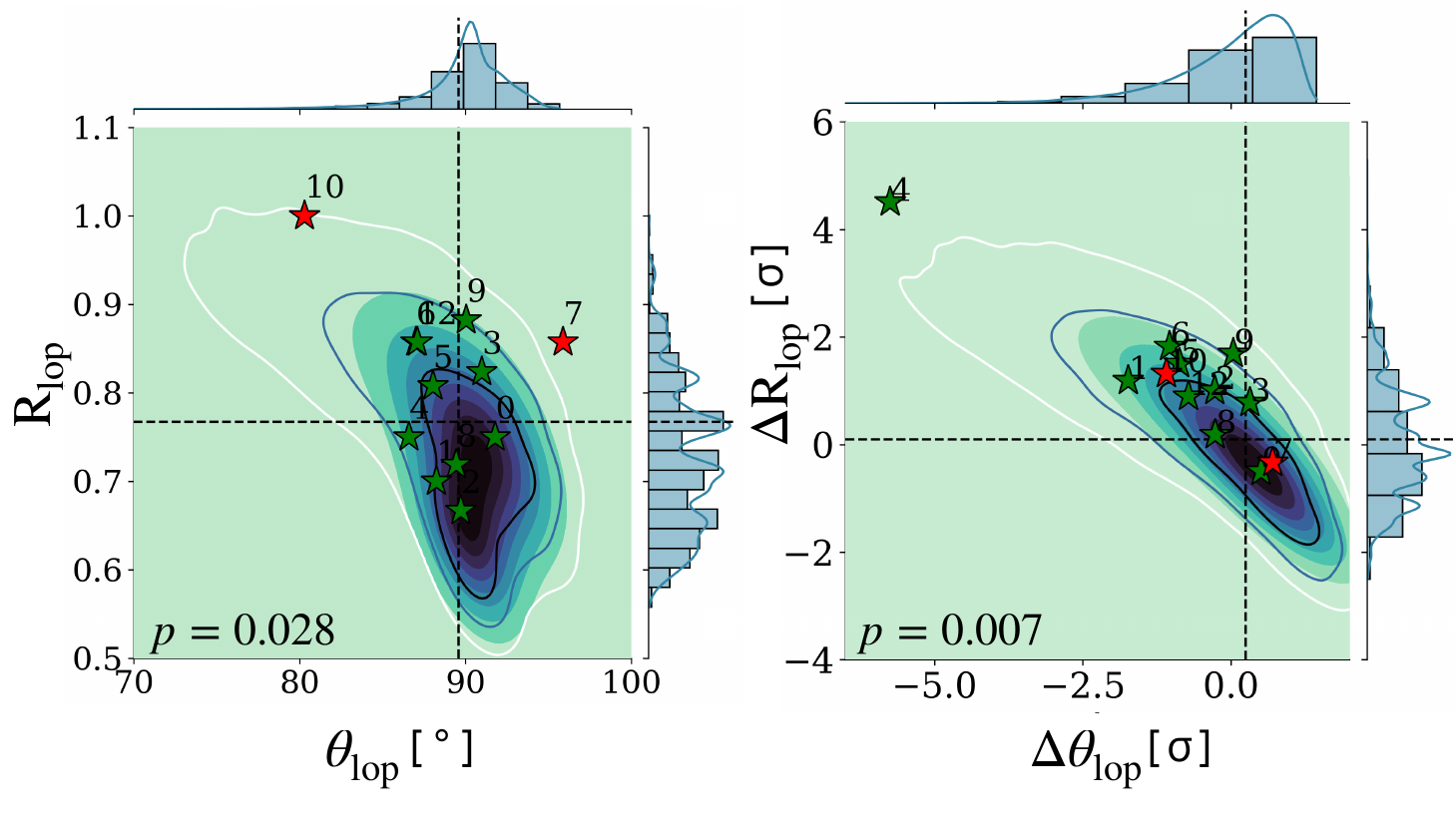}
    \caption{{\it Left:} Joint distribution for $R_{\rm lop}$ and $\theta_{\rm lop}$: (1) in the stacked halo-tracing (null) hypothesis in shades of blue with 1$\sigma$, 2$\sigma$, and 3$\sigma$ solid contours. (2) For real NewHorizon systems as green stars. Systems with strictly less than 10 satellites are indicated as red stars for reference but not included in calculating the underlying joint distribution or in the comparison. {\it Right:} Same analysis for $\Delta R_{\rm lop}$ and $\Delta \theta_{\rm lop}$,the deviations from the means of each system's individual halo-tracing null-hypothesis. Marginal histograms show the corresponding one-dimensional distributions, with dashed lines indicating their means.}
    \label{fig:rlop_vs_thetalop}
\end{figure}

We can then estimate consistency between these two populations in two ways. 
First, we compare directly the full NewHorizon population to the full null-hypothesis population combining all random realizations for all systems. The resulting $R_{\rm lop}$ -- $\theta_{\rm lop}$ joint distributions are presented in Fig.~\ref{fig:rlop_vs_thetalop}, left panel, with the null-hypothesis 1$\sigma$, 2$\sigma$, and 3$\sigma$ contours in shades of blue and real NewHorizon systems as green stars. Removed systems appear as red stars.

It is apparent that the true NewHorizon distribution is shifted towards stronger lopsidedness, especially hemispheric lopsidedness ($R_{\rm lop}$), compared to the mere halo-tracing hypothesis. 
It is confirmed by a Fasano-Franceschini test (FF, 2D extension of KS test) on the two samples, the NewHorizon systems and the null halo-tracing background, which returns a $p-$value$=0.028$. Note that when including Systems 7 and 10 in the analysis ($N_{\rm sat} < 10$), the FF test finds $p= 0.05$, adding noise but still confirming the incompatibility of the two distributions.
Thus, luminous satellites arrange in distributions that are more lopsided compared to mere DM expectations. This distinction is important to consider because the apparent ``unusualness" of systems such as Andromeda is typically assessed against expectations for dark matter sub-haloes (either from pure dark matter or under resolved hydrodynamic runs). This is due to the fact that very few cosmological simulations actually resolve luminous dwarf satellites. But this potentially overstates how anomalous their satellite configurations are relative to the luminous satellite populations predicted by resolved hydrodynamical simulations like NewHorizon.

While informative, this population-level analysis is incomplete as some systems can exhibit strong shifts in lopsidedness compared to what is expected from their DM halo shape only, without particularly standing out in the population. This is in  particular the case of very rich systems in mostly spherical halos, for which even moderate levels of lopsidedness involve large numbers of satellites, hence already represent a strong deviation compared to their individual halo-tracing hypothesis.
To better analyze this aspect, we re-normalize each sample of 5000 random realizations obtained for a given system so as to express the lopsidedness metrics ($R_{\rm lop}$ and $\theta_{\rm lop}$) for each system not in terms of absolute values but in terms of deviations from the mean of the system's {\it individual} halo-tracing null-hypothesis sample. More specifically, we define:
\begin{equation}
  \Delta \theta_{\rm lop} =  \frac{\theta_{\rm lop} - \langle  \theta_{\rm lop}\rangle^{null_{ind}} }{\sigma^{null_{ind}}_{\theta_{\rm lop}}}  ;  \Delta R_{\rm lop} =  \frac{R_{\rm lop} - \langle  R_{\rm lop}\rangle^{null_{ind}} }{\sigma^{null_{ind}}_{R_{\rm lop}}} 
\end{equation}
the relative deviations of each system to its individual ({\it ind}) null-hypothesis average. 

The stacked results and normalized null-hypothesis are presented in Fig.~\ref{fig:rlop_vs_thetalop}, right panel. The incompatibility between the distribution of real systems and the normalized null hypothesis is still obvious and, in fact, strengthened with a FF test returning a p-value$=0.007$. However, now System 4, our richest system $N_{\rm sat}=104$ stands out as particularly lopsided for its halo shape, with a $>5\sigma$ deviation in both $R_{\rm lop}$ and $\theta_{\rm lop}$ from naive DM-based resampling expectation. It should be noted that System 4 is one of the most massive halo in the sample, and linked to System 2, another massive system ($N_{\rm sat}=84$) in our range, by a dense cosmic filament. In total, out of our 12 NewHorizon systems, 1 reaches $>5\sigma$ and 1 reaches $>2\sigma$. While extreme lopsidedness (compared to DM) is therefore not the norm in NewHorizon, it is also not rare. This confirms that luminous satellites in sufficiently resolved hydrodynamic simulations are significantly more lopsided than DM-based expectations. It is therefore unclear that Andromeda remains an anomaly in this context. 

To assess the sensitivity of these results to satellite sampling, we additionally repeat the analysis after sub-sampling sufficiently rich NewHorizon systems to match Andromeda's satellite richness ($N_{\rm sat}=37$). As discussed in Appendix~\ref{appendix:sampling-bias}, the resulting distributions remain inconsistent with the halo-tracing hypothesis, reinforcing our conclusion that the enhanced lopsidedness of luminous satellites cannot be explained by differences in satellite richness alone.

\subsection{Other dark matter inherited lopsidedness?}

So far, we only tested the lopsidedness of NewHorizon systems against their overall DM halo shape. But other sources of lopsidedness can still be the result of dark-matter dynamics. In particular, DM sub-halo clustering and group infall can enhance lopsidedness at least temporarily. More generally, any form of anisotropic accretion of DM can contribute to lopsidedness.

\begin{figure}
\centering
	\includegraphics[width=1.\columnwidth]{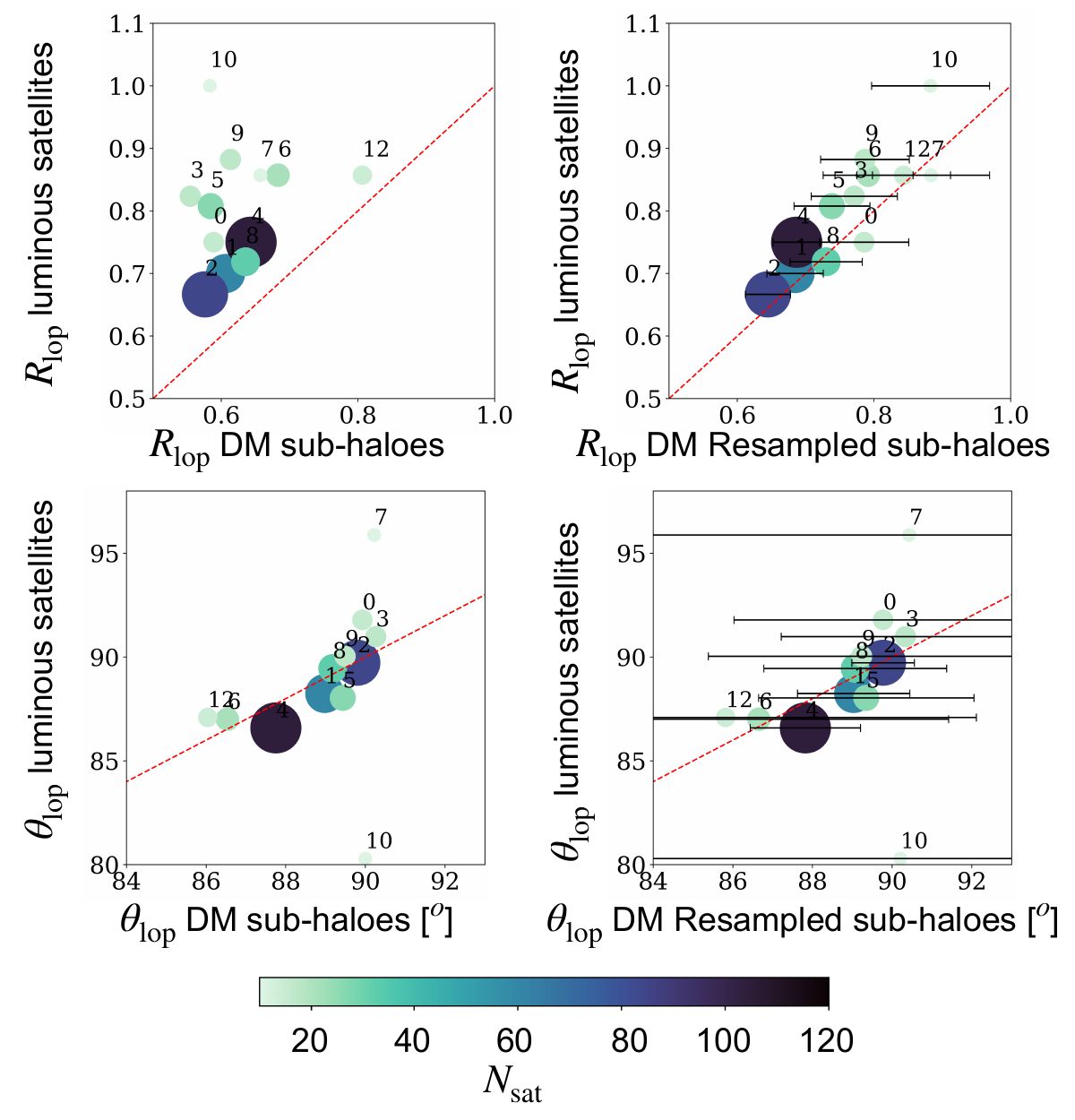}
    \caption{ $R_{\rm lop}$ ratio \textit{(top)} and $\theta_{\rm lop}$ \textit{(bottom)} for the distribution of all luminous satellites vs for all DM sub-haloes within $R_{\rm vir}$ for all 12 MW systems in NewHorizon. The right panel in both rows show the median value of all DM subhaloes' $R_{\rm lop}$ and $\theta_{\rm lop}$ after resampling each system’s DM subhaloes at the number of luminous satellites, $N_{\rm sat}$, of each system. The colors and sizes of each system both correspond to $N_{\rm sat}$. Luminous satellites display more lopsided distributions than their DM counterparts in terms of $R_{\rm lop}$.}
    \label{fig:lum_vs_dm}
\end{figure}

To fully test to which degree the lopsidedness of luminous systems in NewHorizon is inherited from the underlying DM dynamics, Fig.~\ref{fig:lum_vs_dm} displays the two previous metrics of each system, $R_{\rm lop}$ and $\theta_{\rm lop}$, calculated for luminous satellites only vs. for DM sub-haloes in the system's host halo. Left panels show the results for all sub-halos in the host, while right panels show the results when sub-sampling the host halo to $N_{sub}=N_{\rm sat}$, hence allowing to account for sampling biases. The sub-sampling is performed 1000 times so as to display mean and spread for each system.

It is apparent that $\theta_{\rm lop}$ is mostly insensitive to resampling, and consistent across DM and luminous measurements likewise, but with wide variations depending on the specific sub-sample. This is consistent with it being mostly sensitive to general anisotropic accretion. More generally, this metric appears very noisy and not highly sensitive. In particular its purported sensitivity to local clustering is called into question given its lack of sensitivity to resampling on average.

However, the trend for $R_{\rm lop}$ is markedly different. Luminous satellites are systematically biased towards higher $R_{\rm lop}$ (planar-asymmetry lopsidedness), even after correcting for sampling differences, an effect particularly strong for low-to-medium richness systems ($N_{\rm sat}<50$). After correcting for sampling bias, the effect typically reaches a $15\%$ difference between DM and baryons for systems found to be significantly lopsided in this study. This confirms that the marked lopsidedness of luminous systems is dominated by hemispheric asymmetry and is in large part a baryonic effect.
However, the offset but tight relationship between $R_{\rm lop}^{\rm lum}$ and $R_{\rm lop}^{\rm DM}$, on the top-left panel of Figure \ref{fig:lum_vs_dm}, even after correcting for sampling, suggests a feature connected to DM accretion, but significantly amplified by baryon dynamics. This feature may suggest that accretion along cosmic filaments is at play. Indeed, in \cite{Madhani2026Zippers}, we showed that intermediate-scale cosmic filaments, typically connected to Milky Way-mass galaxies, are characterized by spinal gas accretion that is directionally constrained by dark matter accretion but significantly more focal. This direct effect of the dissipative nature of gas is only detectable in cosmological simulations that maintain sufficiently high spatial resolution ($<15$ kpc) in these outer environments, several megaparsecs from MW-mass halos \citep{Pichon11, 2011arXiv1106.0538K,Lu2024,GalarragaEspinosa2022}.In \cite{Madhani2026Zippers}, this collimating effect was sufficient to explain the formation of thin planes of dwarf satellites similar to those seen around the Milky-Way. Diversity among these filaments, and the presence of a dominant one, may therefore explain lopsidedness in NewHorizon too. We explore this topic further in Section.~\ref{sec:cweb}.

\subsection{How focal is lopsidedness in NewHorizon?}

The weak population-level signal in the pairwise-angle distribution, but systematic shift towards higher $R_{\rm lop}$ suggests that the primary baryonic anisotropy present in NewHorizon systems is not small-scale clustering but planar asymmetry. However, satellites may also occupy broader preferential angular regions that are insufficiently captured by pairwise statistics alone. This motivates a cone-based analysis, in which cones of varying opening angles are systematically rotated to identify the directional vector that maximizes satellite number over-density. By comparing these observed maxima, $N_{\theta}$ (one per opening angle), to those obtained from halo-tracing mock realizations, this method (see Section~\ref{sec:methods1}) probes lopsidedness as a function of angular scale and provides a more sensitive test of extended anisotropic configurations. 

\subsubsection{Cones of maximal asymmetry: example systems}

\begin{figure*}
	\includegraphics[width=\linewidth]{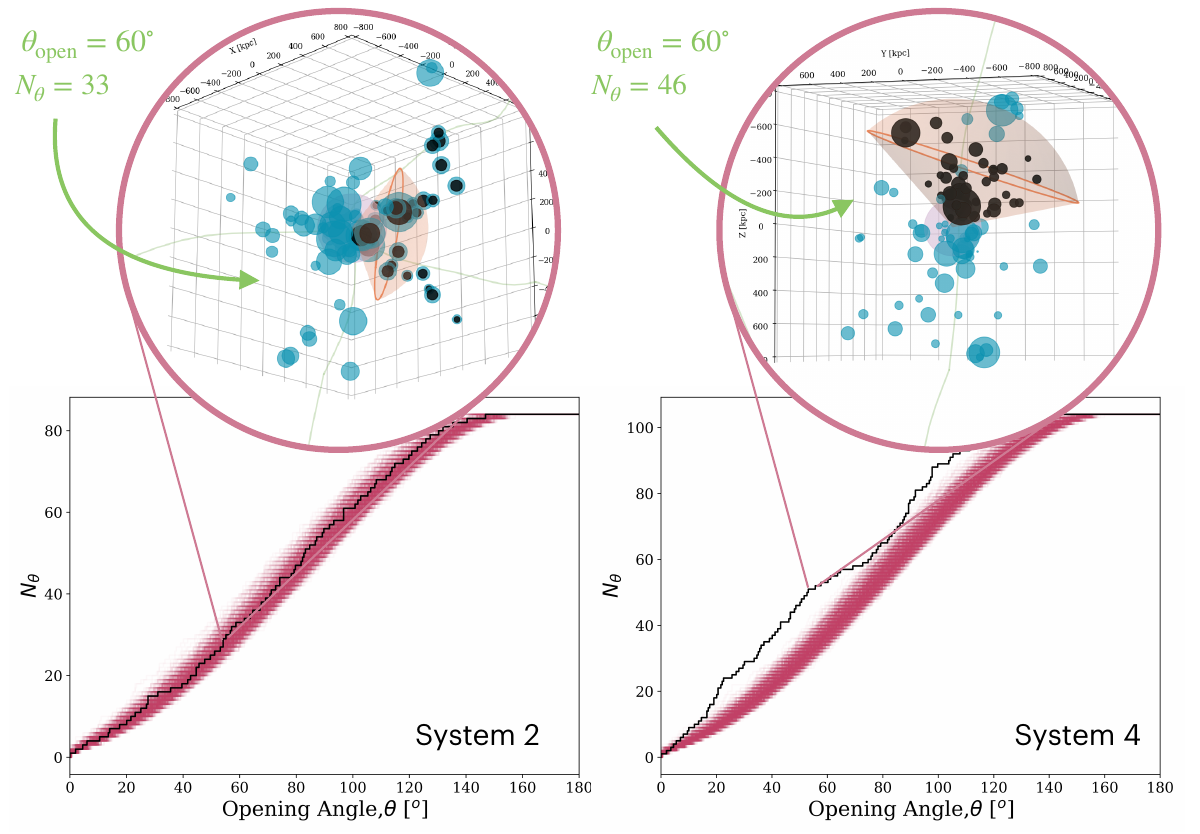}
    \caption{The NewHorizon observed distribution of $N_\theta$ for System 2 with $N_{\rm sat} = 84$ at left and System 4  with $N_{\rm sat} = 104$ at right, are shown in solid black. Distributions of $N_\theta$ for 5000 mock systems drawn from the halo-tracing hypothesis are shown underlying in dark pink. We show an example of the calculation of the satellite count to measure $N_\theta$ in the zoom inserts. In light orange, we show the shell of a cone with opening angle, $\theta_{\rm open}$ of $60^\circ$. The axis of the cone is rotated 1000 of times in order to maximize the orientation that encompasses the greatest number of satellites, $N_\theta$. We show all satellites in light blue and the central galaxy in light purple. In light green lines, we show the skeleton of the underlying cosmic web.  For System 2 at left, $\theta_{\rm open} = 60^\circ$ corresponds to a relative agreement with the null distribution. In contrast, $\theta_{\rm open} = 60^\circ$  for System 4, corresponds to an excess in comparison to the null, indicating that satellites are much more anisotropically located within the halo along a preferred direction, which is consistent with its highly planar structure.}
    \label{fig:cone_bg}
\end{figure*}

In Figure \ref{fig:cone_bg}, we present the $N_\theta$ distribution of two of the NewHorizon systems, Systems 2 and 4, in black. We show the null distribution in dark to light pink -- i.e. $N_\theta$ calculated on a sample of 5000 synthetic distributions per NewHorizon system, each preserving $N_{\rm sat}$ and tracing the underlying DM halo shape. For every cone opening angle, we show the maximum satellite count in the null distribution as well as in the NewHorizon system. In addition, 3D illustrations are presented on the example of $\theta = 60^\circ$. Note that, as the opening angle nears $180^\circ$, or a full sphere, $N_\theta$ approaches $N_{\rm sat}$.

System 2 is one of the least lopsided systems in NewHorizon, which is confirmed by the strong consistency between its $N_{\theta}$ function and its null hypothesis. System 4, a planar system, is on the contrary strongly lopsided, with $N_{\theta}$ past $3\sigma$ deviation across a large range of opening angles $[20^\circ, 140^\circ$, and maximum deviations seen for opening angles $[20^\circ, 60^\circ, 100^\circ]$. This significant excess of $N_\theta$ at lower opening angles ($<180^\circ$) indicates an anisotropic distribution of satellites tighter than pure planar asymmetry, with satellites showing a broad preferred direction. This once again suggests a connection with the cosmic web and coherent filamentary infall.  Indeed, System 2 is our most connected system, largely isotropic and, as expected, at the intersection of multiple dense filaments, while System 4's accretion environment is dominated by the dense filament connecting it to System 2.

\subsubsection{Population analysis in NewHorizon}

\begin{figure}
	\includegraphics[width=1.1\columnwidth]{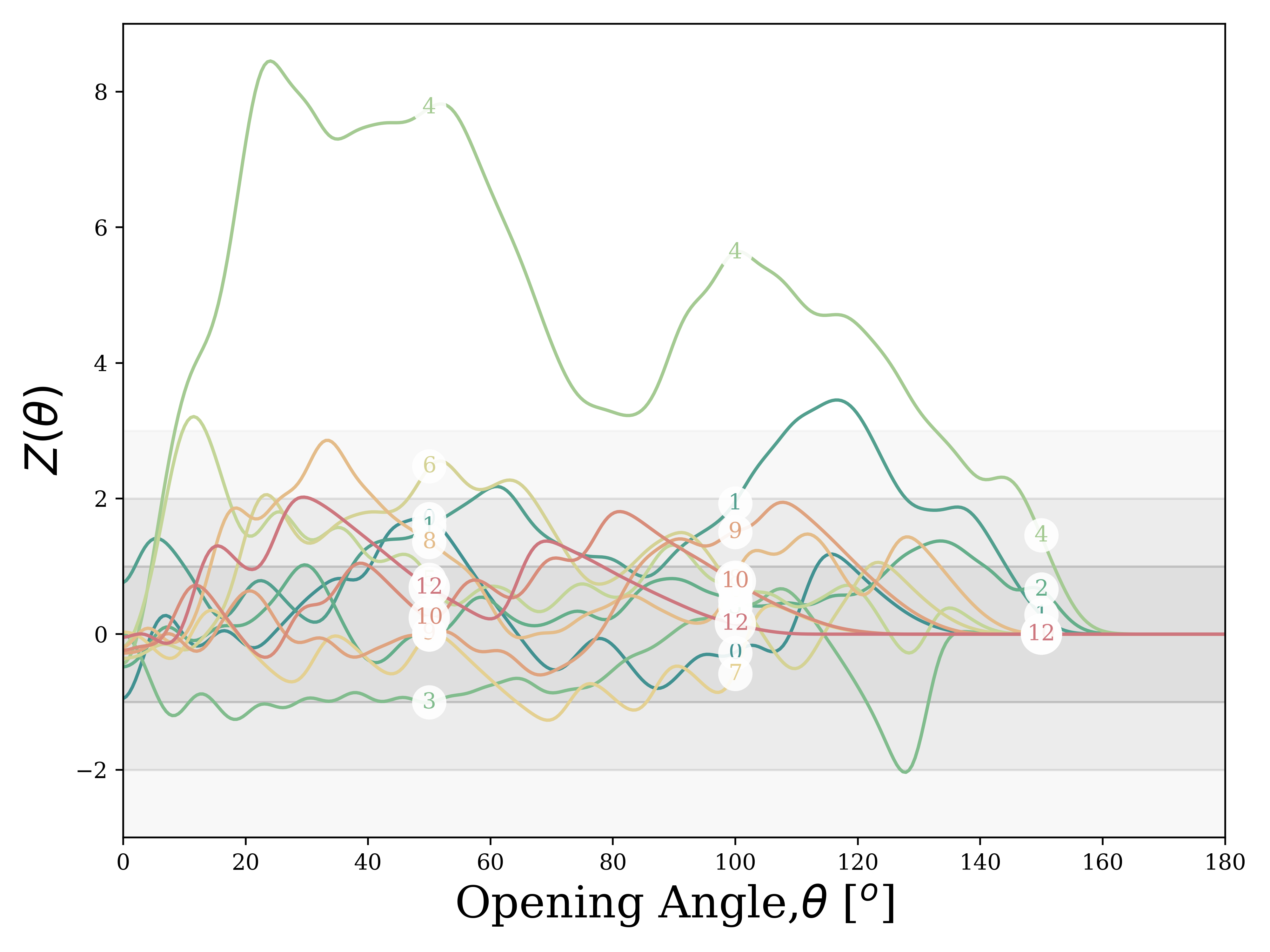}
    \caption{$Z(\theta)$ standard deviation curves for all 12 NewHorizon systems calculated with respect to the mock isotropic background are shown as colored lines and labeled with the System ID. The transparent gray bands in graduating opacity highlight the standard deviation range 0-1$\sigma$, 1$\sigma$-2$\sigma$,and 2$\sigma$-3$\sigma$, respectively. At higher opening angles, the $N_\theta$ of both the null distribution and the NewHorizon distribution converge to $N_{\rm sat}$ resulting in a convergence of $Z(\theta) \rightarrow 0$. }
    \label{fig:ztheta_all_systs}
\end{figure}

To show results for the entire population of NewHorizon, we quantify angular over-densities using the statistic $Z(\theta)$ described in Section \ref{sec:cone_stat}, defined at each cone opening angle $\theta$ as the deviation of the observed maximum satellite count, $N_{\theta}$ from its null expectation, normalized by the standard deviation of the null at that angle. In Figure \ref{fig:ztheta_all_systs}, we present the standard deviation curves, $Z{(\theta)}$ of all 12 NewHorizon systems. We notice that strong angular concentrations ($>2\sigma$) are present in a subset of 5 NewHorizon systems -- Systems 1, 4, 5, 6 and 8 -- and very strong ones ($>3\sigma$) in 3 NewHorizon systems. Hence 40$\%$  of satellite systems show an angular lopsidedness significantly above their halo-tracing expectation, with most maintained over a range of angles of at least $60^\circ$. Although evaluating $Z(\theta)$ across multiple opening angles introduces a multiple-testing concern, the significant deviations persist over a broad angular range rather than appearing as isolated peaks. As such, the global significance of each outstanding system remains above $2\sigma$ under the Simes correction for multiple comparisons.

Most systems with $Z(\theta)>2.5\sigma$ start displaying such strong angular concentrations at opening angles $\theta<90^\circ$ (System 5: $5-20^\circ$, System 8: $20-40^\circ$, System 4: first plateau at $20-60^\circ$, System 6: $55-70^\circ$). This suggests that lopsidedness in NewHorizon is more focal than pure planar asymmetry (i.e. peak at $90^\circ$), or at least embeds a somewhat focal component. However, apart from System 5, 4/5 of these systems still peak at relatively large opening angles $>35^\circ$ not compatible with small-scale clustering either, a result consistent with the relatively small deviations observed for $\theta_{\rm lop}$.

Both System 4 and System 1 also peak at larger angles ($105^\circ$ and $115^\circ$ respectively) with $>3\sigma$ deviations, indicating more hemispheric lopsidedness. These are strikingly compatible with results obtained for M31 in \cite{Kanehisa_2025}, which display a 3-4 $\sigma$ maximal deviation from its DM-derived null-hypothesis at an opening angle of $105-115^\circ$ and large range of angles $>2\sigma$ deviations. This suggests that M31 may not be particularly anomalous compared to $\Lambda CDM$ simulations once these have sufficiently resolved environmental hydrodynamics, like NewHorizon. We explore this further in Appendix~\ref{appendix:m31}. Of note, System 9 appears as a borderline significant system reaching $Z(\theta)>1.9\sigma$ between $107-112^\circ$ (close to hemispheric lopsidedness) but with no focal component. This is in line with it appearing borderline significant ($2\sigma$) in the $R_{\rm lop}$-$\theta_{\rm lop}$ plane while having $\Delta\theta_{\rm lop}\sim 0$.

In summary, most anomalous forms of lopsidedness in NewHorizon seem to remain close to hemispheric $60^\circ-120^\circ$, but a number of anomalous systems include a broad focal component. Narrow focal component is almost absent in our sample, with the exception of System 5 for which it is extreme, and associated with a group infall event. In the following section, we explore the cosmic origin of this ubiquitous lopsidedness.

\section{Merging Filaments as a driver of Lopsidedness}
\label{sec:cweb}

\subsection{Orientation of lopsidedness in cosmic filaments}

\begin{figure*}
    \centering
    \includegraphics[width=\linewidth]{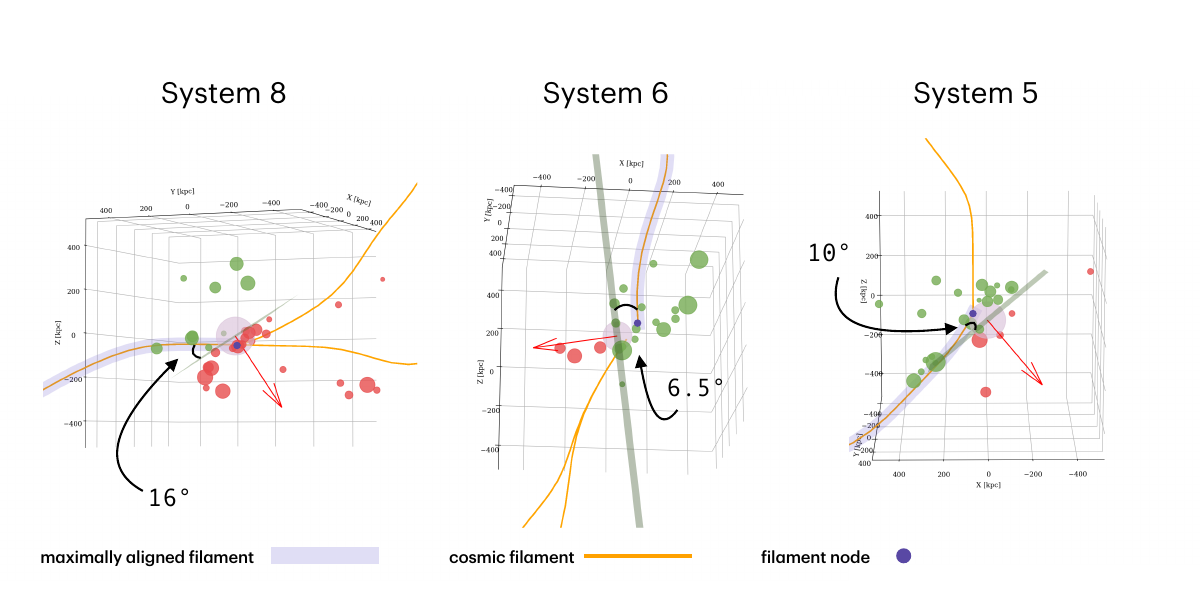} 
    \caption{Examples of the plane of maximum asymmetry identified for three galaxy systems in NewHorizon, Systems 8, 6, and 5 (shown left to right). Satellites appear color coded by red or green depending on which hemisphere they populate as split by the plane of maximum asymmetry, shown in light green seen edge on. The central galaxy appears as light purple. The orange lines show the skeleton of the cosmic filaments identified by \texttt{DisPerSE}. The small dark green circle located at the juncture of multiple filaments depicts a node. The thick light purple band that encloses an orange cosmic filament highlights the individual filament that is maximally aligned with the plane of maximum lopsidedness.}
    \label{fig:maxalign_fil}
\end{figure*}

To investigate the role of cosmic filaments in driving lopsidedness, we first estimate the degree of alignment between the plane of maximal asymmetry and filaments connecting to the halo (characterized by angle $\theta_{\rm cosmic}$). Fig.~\ref{fig:maxalign_fil} shows 3D maps for 3 typical systems, with maximally aligned filament shaded in grey and plane of maximum asymmetry in light green. The rest of the systems are presented in Appendix.~\ref{appendix:alignments}.

\begin{figure}
    \centering
    \includegraphics[width=\linewidth]{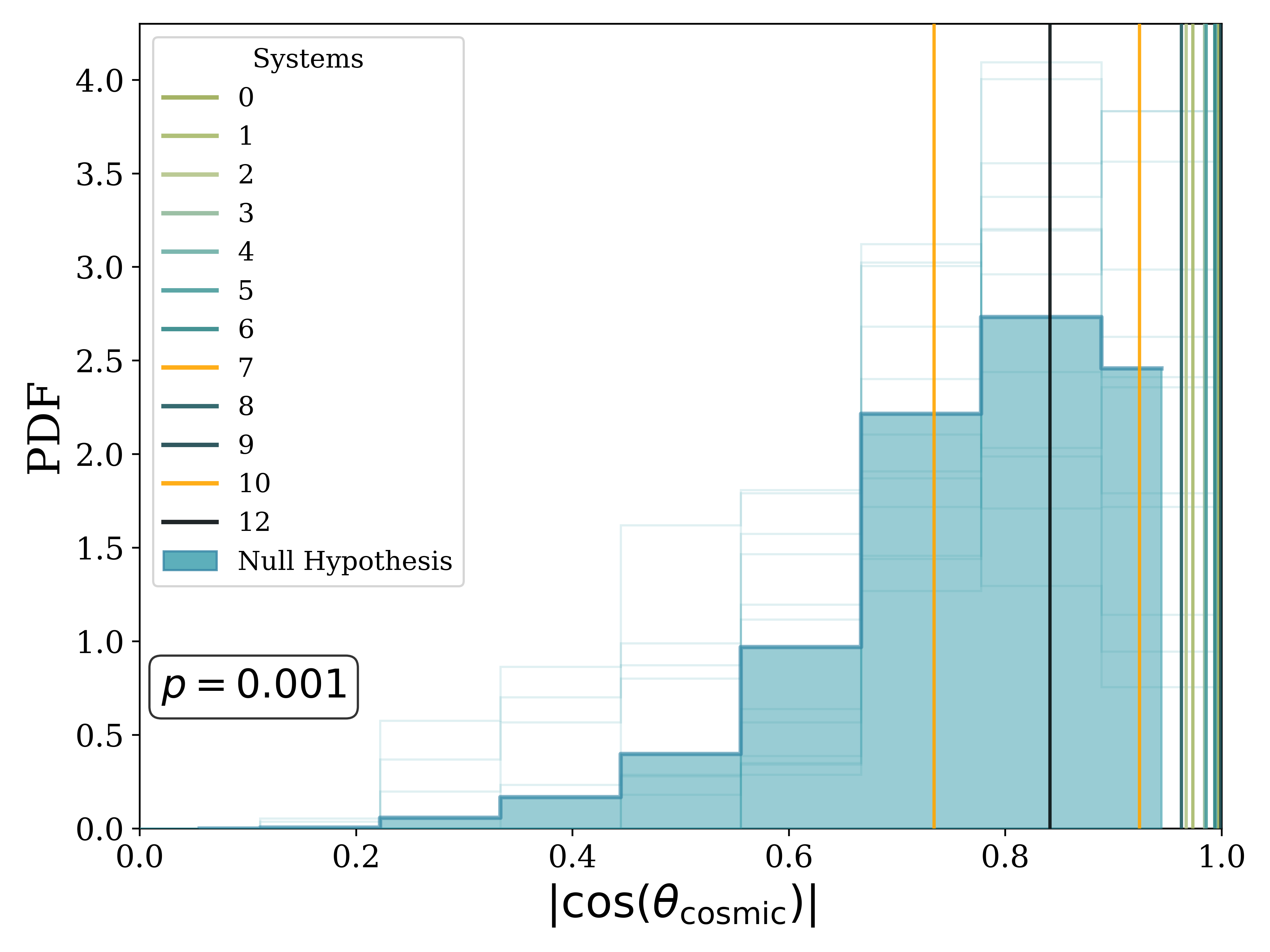} \caption{Stacked null distributions of $|\cos(\theta_{\rm cosmic})|$ for the 12 MW systems in NewHorizon are shown in translucent light blue. The average null distribution is shown as a filled histogram and outlined in blue. Vertical lines indicate the actual NewHorizon $|cos(\theta)|$ values, for $\theta_{\rm cosmic}$ -- the angle between the plane of maximum asymmetry and the maximally aligned nearby cosmic filament of each NewHorizon system. The orange lines demarcate systems with $N_{\rm sat} < 10$, highlighting that the majority, $> 80\%$, of NewHorizon systems are the tail-end of the $|cos(\theta)|$ distribution and highly aligned with nearby cosmic filaments. The $p$-value of the Anderson-Darling test of the observed values with the null distribution finds $p= 0.001$.}
    \label{fig:costheta_filalign}
\end{figure}

It is striking that planes of maximal lopsidedness are almost systematically aligned with a branching cosmic filament, with an average filament-plane angle of $13.9^\circ$ across the whole sample, and an average of  $10.2^\circ$ across the 5 most anomalously lopsided systems as identified in Fig.~\ref{fig:ztheta_all_systs}. To quantify the significance of this behavior, in Fig.~\ref{fig:costheta_filalign}, we compare the distribution of angles $\theta_{\rm cosmic}$ across the 12 NewHorizon systems (vertical lines) against a composite null-hypothesis in which each system, characterized by its fixed set of branching filaments, is equally represented, and $\vec{n}_{\rm lop}$ resampled 5000 times per system (blue-shaded histogram). 
While some degree of alignment is expected due to the fact that most systems are connected to more than 1 filament, hence have more than one candidate for alignment, it is visible on Fig.~\ref{fig:costheta_filalign} that the measured degree of alignment of lopsidedness planes with filaments is much higher than predicted by chance (p-value $<0.001$).

As discussed in Methods, due to the small number of satellites in some systems the orientation of the plane of maximal asymmetry is determined only up to a certain angular precision, with remains $<10^\circ$ for the vast majority of systems but can reach $15-20^\circ$ for System 7 and 10. However, given the strong alignment of all planes with filaments, these differences only amount to variations of $\Delta(\cos\theta_{\rm cosmic})<0.015$ for all systems but 7 and 10, not modifying the p-value. Moreover, when choosing the average  over all possible  orientations for Systems 7 and 10, their alignment to cosmic filaments is in fact significantly increased, as  detailed in Appendix.~\ref{appendix:degeneracy}.

On the contrary, alignment of the nearest massive neighbor of each system in its plane of maximum asymmetry is detectable but much weaker (see  $\theta_{\rm NN,plane}$ in Table~\ref{table:fil_align}), with a mean angle of $20^\circ$ across the sample, reaching an underwhelming $38^\circ$ across the five most anomalously lopsided systems. In fact, while there is an alignment trend, it barely reaches statistical significance over the null-hypothesis that plane-neighbor alignment is random (KS-test p-value$=0.05$). Of note, no tendency of lopsidedness to orient towards the nearest massive neighbor is found (this would correspond to the NN orthogonal to the plane). Similarly, no significant alignment is found between the direction of the NN and that of the cone of maximal lopsidedness $Z_{\rm max}$ (KS-test p-value$>0.25$).

\begin{table}
\centering

\small
\setlength{\tabcolsep}{3pt}

\begin{tabular}{@{}cccccc@{}}
\hline \hline
System & $\theta_{\rm cosmic}$ [$^\circ$] & $N_{\rm fil}$ & $d_{\rm NN}$ [Mpc] & $\theta_{\rm NN,fil}$ [$^\circ$]  & $\theta_{\rm NN, plane}$ [$^\circ$]\\
\hline
0 & 4.38 & 1 & 3.52 & 167.51 & 1.99 \\
1 & 13.24 & 3 & 2.16 & 143.86 & 36.51\\
2 & 14.71 & 4 & 1.90 & 128.73 & 34.98\\
3 & 10.22 & 2 & 2.42 & 37.97 & 37.43\\
4 & 6.01 & 2 & 1.90 & 169.59 & 9.33 \\
5 & 9.72 & 2 & 3.81 & 156.74 & 6.4\\
6 & 6.49 & 2 & 0.86 & 16.51 & 14.52 \\
7 & 22.49 & 1 & 1.48 & 160.92 & 16.19\\
8 & 15.65 & 3 & 0.84 & 12.34 & 49.69\\
9 & 2.10 & 1 & 0.07 & 119.81 & 25.61 \\
10 & 42.75 & 1 & 0.70 & 29.64 & 12.39\\
12 & 32.73 & 3 & 0.71 & 102.17 & 2.58\\
\hline
\end{tabular}
\caption{Alignments with nearby cosmic filaments are presented here. $\theta_{\rm cosmic}$ is the angle between the maximally aligned nearby filament and the plane of maximum asymmetry.
$N_{\rm fil}$ denotes the host's filament connectivity, defined as the number of filaments connected to its nearest node.
$d_{\rm NN}$ is the distance from the central galaxy to its nearest massive neighboring galaxy.
$\theta_{\rm NN,fil}$ measures the angle between the nearest-neighbor separation vector, $\hat{r}_{\rm NN}$, and the filament axis most closely aligned with the plane of maximal asymmetry, while $\theta_{\rm NN,plane}$ measures the angle between $\hat{r}_{\rm NN}$ and the plane itself.}
\label{table:fil_align}
\end{table}

A closer analysis of the distances of NewHorizon systems to NN further reveals that the NN is more than 1.9 Mpc,  away (or approximately 5-10 $R_{\rm vir}$ for a typical MW mass to Cen A mass halo) for 3/5 of the most anomalously lopsided systems. It is however at distances suggesting stronger interactions ($<900$ kpc) for the two remaining systems. As a reminder, virial radii for our systems typically range from 200 to 330 kpc, with only the three most massive systems between 350 and 450 kpc. Note that the the NN is selected to identify massive neighboring galaxies potentially capable of exerting significant tidal perturbations on the satellite system's geometry, hence boost lopsidedness. This does not preclude the existence of more minor interactions at closer distances but this is not the topic of this study.

Direction towards the nearest neighbor is however itself significantly aligned to the filament most aligned with the plane of lopsidedness (see Table~\ref{table:fil_align}, KS-test p-value$=0.003$). However it should be noted that it is frequently found "overhsooting", i.e on the opposite side of, the system compared to the filament (8/12), ($\theta_{\rm NN,fil}>90^\circ$ in Table~\ref{table:fil_align}). This may indicate that some of these pairs are past the first passage in their interaction. However Fig.~\ref{fig:maxalign_fil} and Fig.~\ref{fig:allfil} also reveal that it is often because many systems are better described as being in the middle of one dominant filament in formation (i.e. undergoing a filament merger). 

Overall, these results suggest that the main alignment of the plane of maximal asymmetry is with the cosmic filament, while alignment with a massive neighbor is merely coincidental, a consequence of the fact that massive neighbors tend to be channeled by filaments in the first place.


\subsection{Hemispheric lopsidedness is anti-correlated to connectivity}
\label{sec:hemisphere}

\begin{figure}
	\includegraphics[width=\columnwidth]{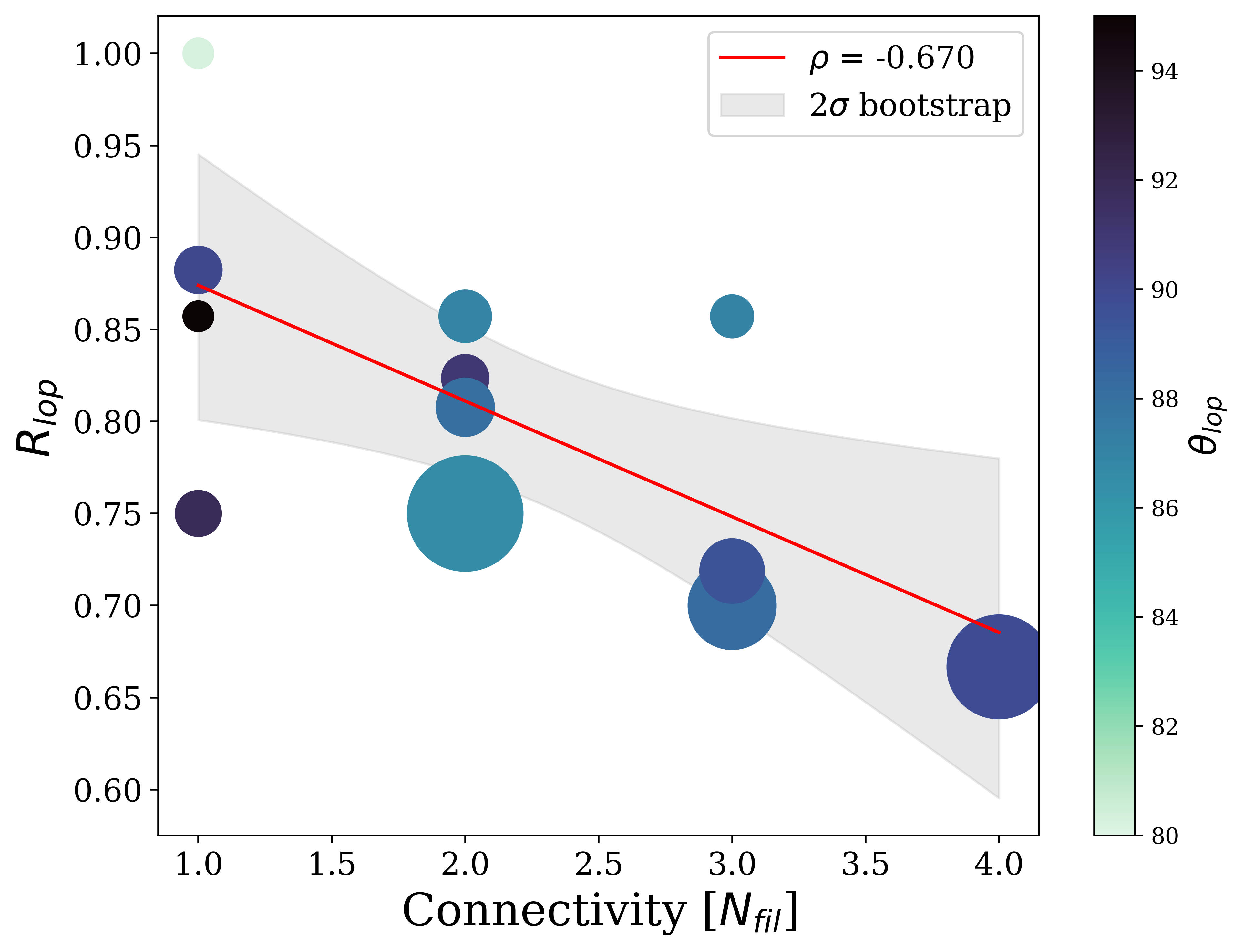}
    \caption{Hemispheric lopsidedness fraction vs. cosmic filament connectivity for NewHorizon systems. Best-fit is displayed on red and 2$\sigma$ bootstrap uncertainties in grey. Dot size scales with $N_{\rm sat}$ and color with $\theta_{\rm lop}$. The lopsidedness fraction is inversely correlated to connectivity. }
    \label{fig:Rlopvsconnectivity}
\end{figure}

To further investigate how filaments may drive hemispheric (``planar") lopsidedness in satellite distributions, let's further analyze the evolution of $R_{\rm lop}$ with connectivity, i.e the number of filaments entering the host halo ($R_{\rm vir}$) of each system. Fig.~\ref{fig:Rlopvsconnectivity} displays the trend for the 12 NewHorizon systems, with dot size fitted to $N_{\rm sat}$ and a Pearson linear regression coefficient listed as $\rho=-0.67$.

The degree of hemispheric lopsidedness $R_{\rm lop}$ is clearly anti-correlated to cosmic connectivity, with most highly lopsided systems ($R_{\rm lop}>85\%$) showing $N_{\rm fil}\leq 2$ while all non-lopsided systems ($R_{\rm lop}<70\%$) display $N_{\rm fil}\geq 3$.
It should be noted that the result still holds when using $\Delta R_{\rm lop}$ instead of $R_{\rm lop}$ to better account for differences in sampling. In this case, the linear regression yields coefficient $\rho=-0.76$.
Eventually, although sampling is limited in this latter case, the analysis also holds when restricted to the three most massive systems (1,2,4) that can be resampled to $N_{\rm sat}=37$, for which the trend $\Delta R_{\rm lop} \propto -N_{\rm fil}$ becomes exact.

The latest result is not surprising as a connectivity $N_{\rm fil}\geq 3$ implies a number of misaligned accretion channels supplying satellites along multiple directions, hence limiting broad angular anisotropies. The picture that emerges is therefore one where filaments provide the main channels of accretion of satellites into halos, hence drive some degree of lopsidedness, providing that some mechanism heavily constrain the directions of accretion to one main hemisphere or direction. This is developed in the next section.

\subsection{Role of filament geometry and dynamics in driving lopsidedness}
\label{sec:zip}

At first glance, the role of of the most aligned filament may still appear counter-intuitive since hemispheric lopsidedness in NewHorizon is arranged such that said filament is aligned with the dividing plane between the well-populated and low-populated sides, {\bf not} towards the most populated side. However, this discrepancy is resolved when analyzing the spatial configurations of our systems in Fig.~\ref{fig:allfil} and Fig.~\ref{fig:maxalign_fil}. Lopsidedness in NewHorizon systems corresponds to two distinct geometric situations, described in the next subsections.

\subsubsection{Nodal systems}

 For nodal systems (1,2,3,4,5,8,12) -- systems with a cosmic node identified $<  R_{\rm vir}$ and $N_{\rm fil}\geq 2$ -- the plane of maximum asymmetry simply appears as the best dividing plane between one side that contains all or most filaments and one side that contains 0 or 1 filament. In summary, it is the dividing plane that maximizes the contrast between the two hemispheres in terms of number/strength of accretion channels. 4/5 of the anomalously lopsided systems in Fig.~\ref{fig:ztheta_all_systs} fall in this category (1,4,5,8). In particular, in cases where $N_{\rm fil}=2$, the small spatial offset between the system itself and the filament network ( $\simeq 100$ kpc for System 3 and System 5, $\simeq 150$ kpc for System 4) appears sufficient to shield one hemisphere from the main filamentary accretion channels, hence boosting lopsidedness and explaining the alignment of the dividing plane with the filament.
 
\begin{figure*}
	\includegraphics[width=\linewidth]{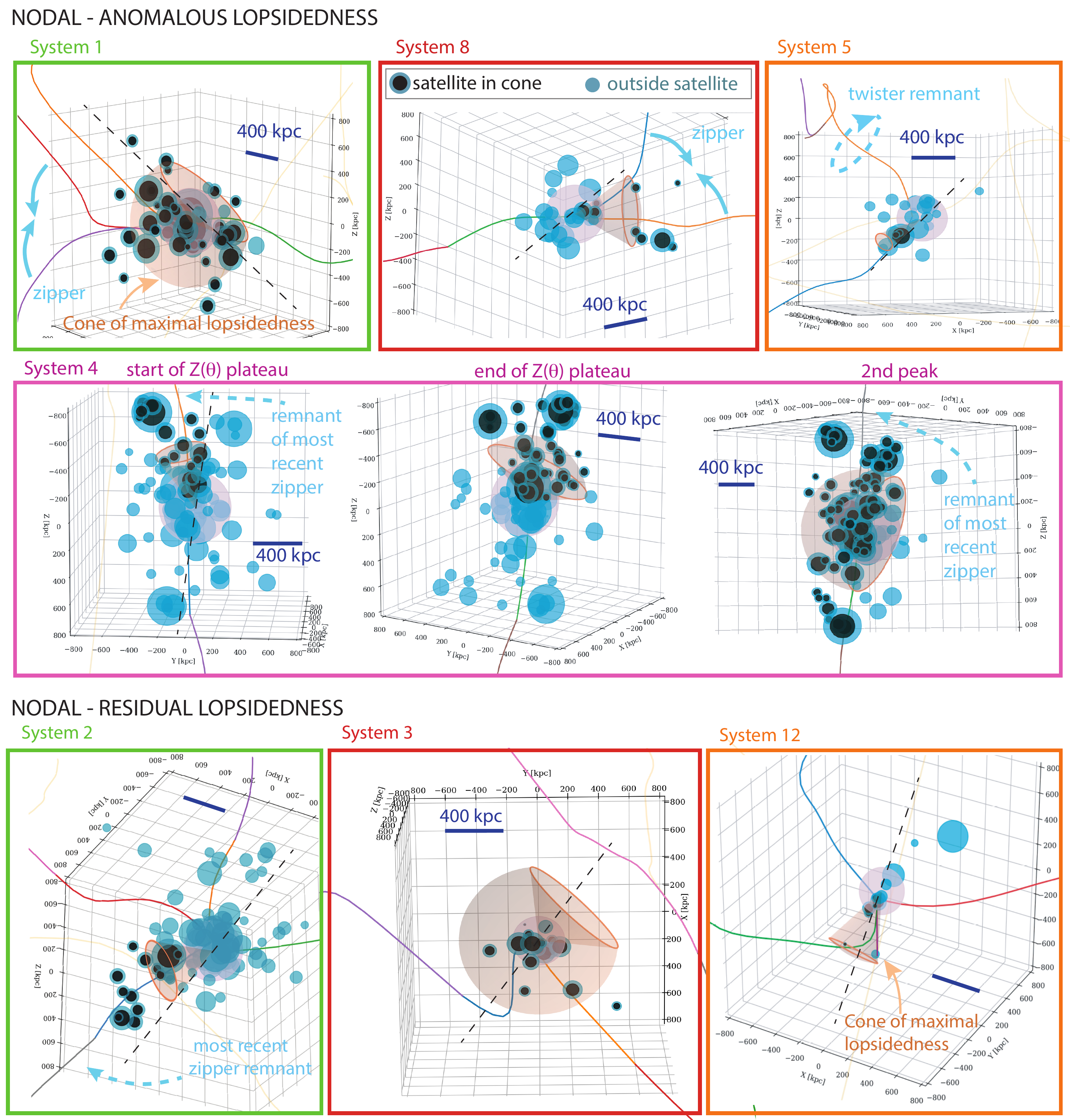}
    \caption{3D maps of cosmic filaments, satellites and cones of maximal lopsidedness around the nodal systems in NewHorizon, with anomalous lopsidedness cases in the upper section. Zippers and twisters (filament mergers with and without orbital momentum) appear as blue arrows. The intersection of the plane of maximum asymmetry with the projection plane centered on the central galaxy is overlaid as dashed line for reference. Note that for System 3 it is actually nearly perpendicular to the line-of-sight, right at the tip of the filaments. Cones of maximal lopsidedness are systematically oriented towards a least one filament, most often encompassing several of them in the process of merging. System 5 is unique in that, while oriented towards a filament, its narrow cone (hence highly focal lopsidedness) is related to a single group infall event.}
    \label{fig:cones1}
\end{figure*}

 Note that this filamentary origin is not limited to hemispheric lopsidedness. For each nodal system, the angle and orientation of the cone of maximal lopsidedness (or the 3 cones corresponding to the 3 highly significant peaks of $Z(\theta)$ for System 4) are also directly related to the angular separation and orientation of the filaments, as can be seen on Fig.~\ref{fig:cones1}, which display filaments, cones of maximal lopsidedness and projected maximal lopsidedness planes for all nodal systems in NewHorizon.
 
Narrow cones (i.e focal lopsidedness)  typically center on 1 filament (Systems 2,4,5), or envelop 2 in the process of merging (Systems 1,8,12) hence with low angular separation with one another at close distance from the system. This applies to both systems with significant lopsidedness and those with residual lopsidedness, as displayed in Fig.~\ref{fig:cones1}. Note that only System 5's very narrow cone is compatible with a single group infall event (but still centered around a filament). 

Broad cones (hemispheric lopsidedness) on the other hand correspond to systems for which the angular distribution of connected filaments departs more significantly from isotropy with, in fact, an arrangement of filaments nearly confined to a single hemisphere (Systems 1, 3, 4). As we have already mentioned, in these cases the small spatial offset between the system's centre of mass and the filaments determines which hemisphere receives most of the accretion (the side with the filament) and which one is deprived of it (the side off the filament). 

In summary, in all cases the cone of max lopsidedness of nodal systems appears directly inherited from the lopsidedness and/or planarity of the global filament network around them.
This is confirmed by the analysis of gas density radial profiles of all filaments surrounding nodal systems. We obtain these profiles for each filament around the system by first projecting the NewHorizon gas density field onto a 3D regular grid with spaxel extent $\Delta x= 14$ kpc, then computing each spaxel's cylindrical (i.e.~tangential) distance to its nearest filament as in \cite{Madhani2026Zippers}. 

In all cases but System 8 (see next section), the densest and cuspiest (in terms of transverse density profile) filaments are found on the maximally populated side of the plane of maximum asymmetry, and cones of maximal lopsidedness are centered on the densest one. Of note, with NewHorizon resolution, the average density of the diffuse gas along the spine  of such filaments frequently reaches 50-100 times the average density of the Universe, a value that drops by a factor 10 within 300 kpc of the densest filament spine for most systems, and within 500 kpc of the spine for more massive filaments branching to most massive systems. This illustrates how narrow accretion channels are into each system and how a relatively mild offset from filaments can deprive an hemisphere of the halo from steady influx of satellites.

\subsubsection{The case of zippers}

 In all studied systems, filaments in the process of merging or recently merged (zippers) tend to concentrate more filaments in a narrow solid angle hence play a role in filament lopsidedness. Indeed, on-going and recent zippers are systematically found on the most lopsided side of the system and within cones of maximum lopsidedness. In addition, still ongoing zippers concentrate filaments without having yet triggered a significant decrease in the number of their satellites through galaxy mergers, which may enhance their contribution to lopsidedness.

\begin{figure*}
	\includegraphics[width=\linewidth]{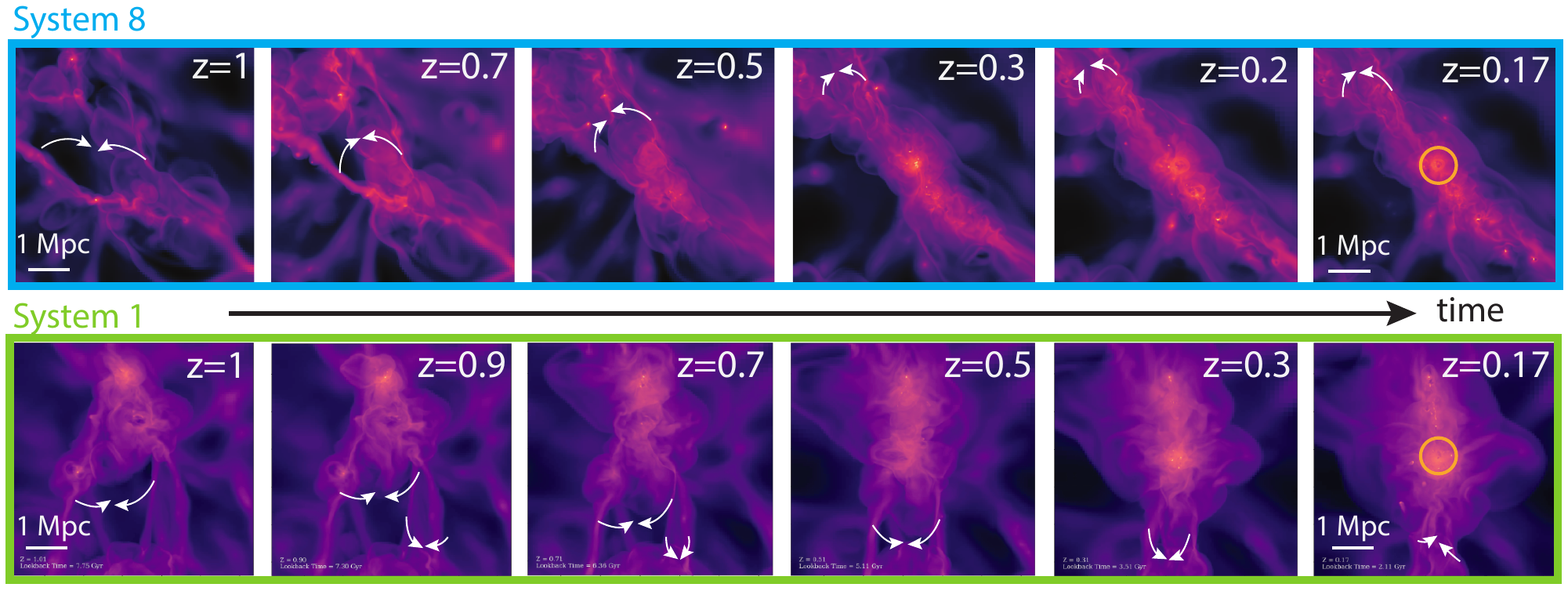}
    \caption{Evolution of the  gas density maps across cosmic time in a fixed co-moving volume around Systems 8 (upper panels) and 1 (lower panels) respectively. On-going Zippers (merging filaments) are shown as white arrows and lead to strong alignment with lopsidedness.}
    \label{fig:zippers}
\end{figure*}

For instance, Fig.~\ref{fig:zippers} shows the progressive zipping of filaments at $z <1$, still on-going at time of observation, surrounding Systems 1 and 8 \footnote{Videos illustrating the evolution of gaseous filaments, including their merging and twisting, in our simulated volume are available at: https://janvimadhani.github.io/filaments}. In particular, System 8 lies at the junction of a mostly straight, on-going zipper of two filaments. In this case, the cone of maximal lopsidedness is directed towards the two strands still in the process of merging (as is always the case when an on-going or recent zipper is present in our sample) rather than the merged side (see. Fig.~\ref{fig:cones1}) {\it even though} these two strands display a spine density $\sim$ twice lower than the merged side. Their number and narrow separation angle however result in their combination providing a more active accretion channel overall, less subject to pre-entry satellite galaxy mergers than the already merged side, hence retaining higher pooling of satellites. It can indeed be seen on Fig.~\ref{fig:maxalign_fil} and Fig.~\ref{fig:cones1} that an substantial part of System 8's lopsidedness can be attributed to the pooling of very low mass satellites in the non-merged section of the filament network.

Let's note that this mechanism is also at play in driving the lopsidedness of System 4 and the residual lopsidedness of System 2 as both narrow cones center on the latest zipper. Although both terminated more than 2 Gyr in the past, these events still boost the satellite density of one filament compared to others.

\subsubsection{Non-nodal systems}

Non-nodal systems (0,6,7,9,10) are skirting one single, locally straight filament, often with a strong side stream, orthogonal with the main filament at large distances but fully realigning with it outside $R_{\rm vir}$ of the system (Systems 6, 7, 9, 10) as can be seen on Fig.~\ref{fig:non-nodal}. In such a case the bifurcation point (smooth junction between filaments) is always $>R_{\rm vir}$. It should be emphasized that, while \texttt{DisPerSe} filament junctions at $z=0.2$ may appear, at first glance, geometrically similar to the ongoing, highly evolutive zippers between filaments around System 8 and System 1 (depicted in Fig.~\ref{fig:zippers}), these cases display in fact markedly different dynamics. In non-nodal systems, the side stream is a typically lower density core density (by a factor $>10$) filament that is not involved in a zipper with the dominant filament, but is in fact stable in shape and direction over the last 5 Gyrs. In particular, it remains orthogonal to the dominant filament at large distances ($>1$ Mpc) across billions of years  \citep[see:][for an example on System 9]{Madhani2026Zippers}.

In each of these cases, the plane of maxim asymmetry is strikingly aligned with the main filament, especially locally. Narrow cones of maximal lopsidedness typically enclose one filament (System 6, 10), one direction being boosted by recent group infall along it, with the exception of System 0 for which the group infall event came from a side stream below the persistence level of our filament detection. Broad cones (System 7, System 9) encompass both sides of their main filament.  In this case, lopsidedness seems to be directly related to the accretion history along the filament or the side stream, depending on whether the system is embedded in one or the other.

System 6 and 9 for instance, are markedly offset from their side stream and seem to mostly accrete from their dominant filament, which latest zipper is $>2$ Gyr in the past. Time evolution analysis also shows that the side stream has remained stable in orientation, away from the system, and has not undergone any zipper since $z\simeq 2$, hence has had minimal influence on satellite funneling into the system across cosmic time. Note that, for System 9, the entire satellite distribution is offset by $< 50$ kpc from the dominant filament but it still maintains near perfect alignment with it. This suggests small drifts of the spine of the filament position over time, potentially limiting recent accretion. Such minor swaying and drifting of filaments around an average position is an expected tidal feature at this scale and has been extensively described in simulations \citep{Pichon11, Jhee26}.

\subsubsection{The case of twisters}

\begin{figure*}
	\includegraphics[width=\linewidth]{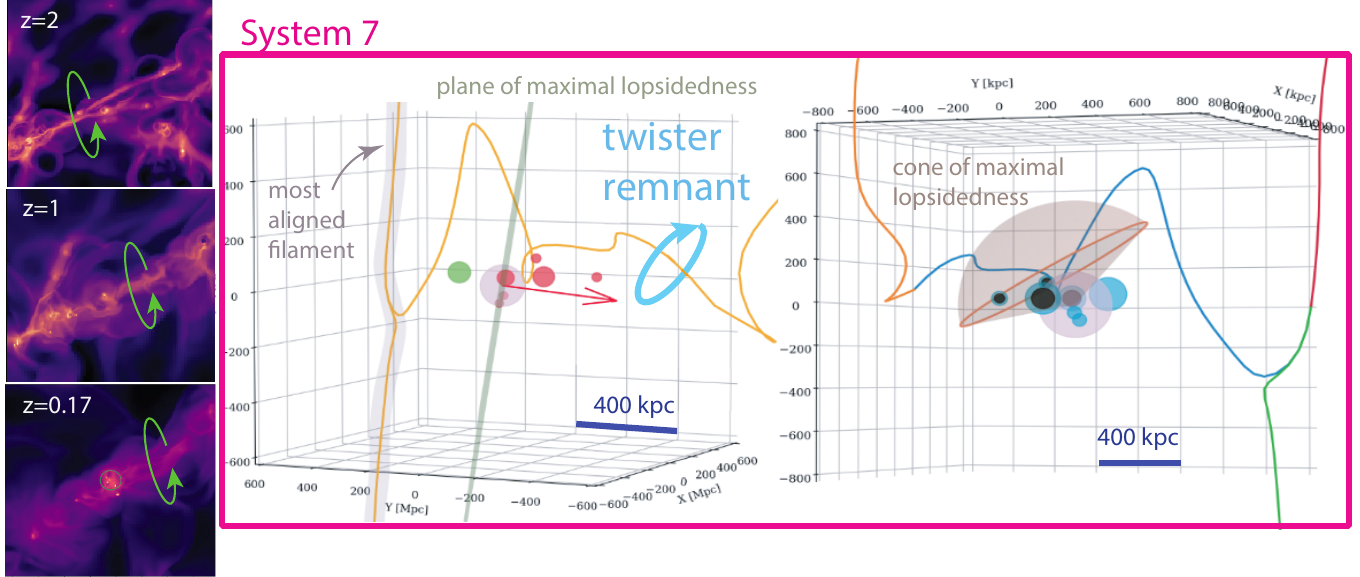}
    \caption{{\it Left}: Gas density maps across cosmic time showing the twister embedding System 7, with main twisting direction shown as a green arrow. {\it Right}: Filaments around System 7 at $z=0.17$, with central/satellite positions and plane and cone of maximal lopsidedness, shown in  yellow as connected segments at left and in multicolor individual segments at right. The panel on the right is rotated in order to help visualize the extent of the twisting filament.  }
    \label{fig:sys7}
\end{figure*}

Filament distortion over time can however be a lot more drastic. The most extreme such case is System 7, which is offset by the dominant filament by $> 300$ kpc and is in fact fully embedded in the orthogonal side stream, as presented on Fig.~\ref{fig:sys7}. It is striking that the \texttt{DisPerSe} extracted spine of the side stream displays high curvature and helicity with loops along its average direction. This curious structure is explained in our previous study \citep{Madhani2026Zippers}. It is a consequence of the history of the stream/filament that recently underwent a twister: a zipper between two filaments characterized by high orbital momentum between the pair of filaments, as illustrated on the left part of Fig.~\ref{fig:sys7}. In such a filament merging scenario, filaments are brought within a narrow separation angle while the high orbital momentum  prevents the formation of a concentrated core, which may limits subsequent galaxy mergers in the remnant. Unsurprisingly, in such a situation the residual lopsidedness is oriented towards the spine of the twister remnant (the main accretion channel) and centered on the highest-helicity region. Indeed the high-vorticity environment (high ``environmental" angular momentum) that this geometry is reminiscent of is already suspected of stalling accretion onto the central and mergers between satellites \citep{Song21, Jhee26}, possibly resulting in a pooling of satellites in that region. The pooling side of System 7 is indeed the one hosting the smallest mass satellites.

While System 7, being fully embedded in the most recent twister, is the most striking example of this phenomenon in our sample, a similar mechanism may be at play in boosting the hemispheric lopsidedness of Systems 10, 0, 6 and even some nodal systems like System 5. In all of these cases, the richest hemisphere is oriented towards the nearby twister remnant (see Fig.~\ref{fig:non-nodal}), while such remnants typically have lower spine densities than other filaments in the network (by a factor 5 to 10) 

\subsection{Summary}

\begin{figure*}
    \includegraphics[width=\linewidth]{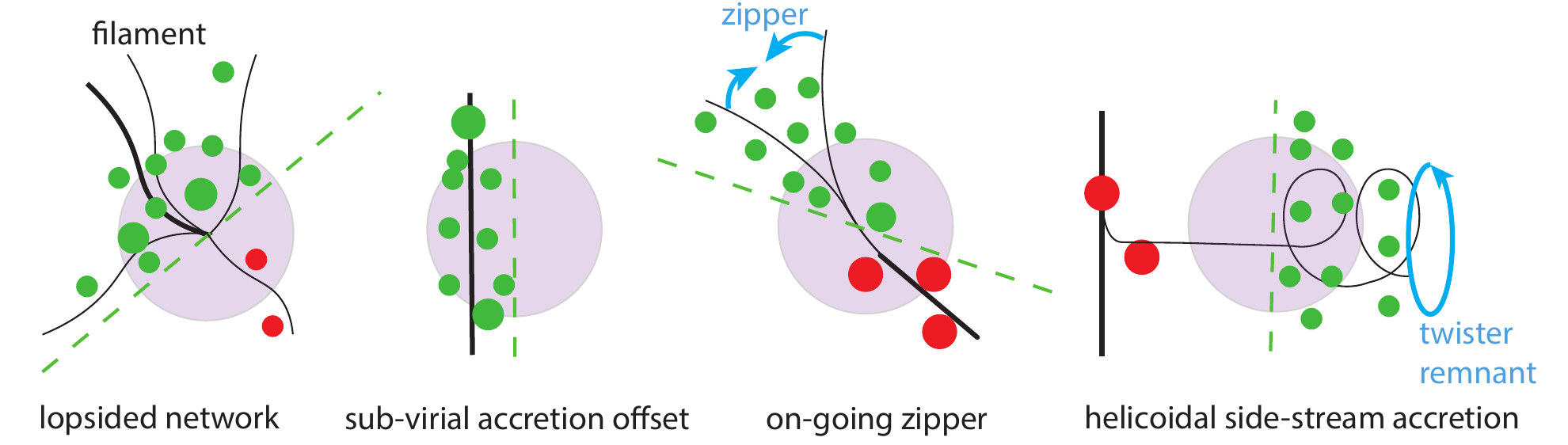}
    \caption{Sketches of the main types of cosmic web-inherited lopsidedness found in NewHorizon. This include purely geometry-driven lopsidedness (lopsided network, filament offset) but also configurations that appear to also be dynamics-driven (zippers, twisters), with less collapsed sections of the network having undergone less galaxy mergers. Different processes can combine.}
    \label{fig:summary}
\end{figure*}

In conclusion, lopsidedness in NewHorizon emerges as a relic of un-relaxed accretion from the highly evolutive, intermediate-scale cosmic web described in this study (i.e~the network of filaments typically connecting halos of $>10^{12-12.5}\, \rm M_{\odot}$). The evolution of this cosmic network is complex, with frequent filament mergers down to the last Gyrs characterized by different amounts of orbital momentum. This is not unexpected since these filamentary scales ($<10$ Mpc) remain far from the scale of homogeneity of the Universe ($> 200$ Mpc), hence are also under the influence of gravitational tides from their anisotropic larger-scale environment. These processes typically result in very lopsided or asymmetric systems of filaments, hence accretion channels around Milk-Way mass and massive galaxies. Our study finds that the lopsidedness of satellites around such galaxies is merely a consequence of this cosmic geometry. Fig.~\ref{fig:summary} summarizes the types of cosmic web lopsidedness found in NewHorizon around Milky-Way mass (and above) systems.

Systems with strikingly anomalous lopsidedness (1,4,8,6) display several aspects of filament evolution and accretion along filaments at once: mostly planar filament network arrangement, on-going or recent zippers boosting certain directions of accretion, recent infall events from same filaments (including massive nearby neighbor) that combine into striking lopsidedness. Unsurprisingly, cone analysis reveals that lopsidedness in NewHorizon is typically broad, extending over $\ge 40^\circ$ solid angle in the case of a recent zipper merger to $>180^\circ$ in the case of a planar arrangement of dominant filaments, with both effects often at play in nodal systems. Residual narrow lopsidedness ($\le 30^\circ$) seen in some systems (5,10,12) is also oriented towards a filament but more typically associated with a recent group infall/merger event, and rarely significant on its own.

As we demonstrated in \cite{Madhani2026Zippers}, resolving both the tides and dissipative hydrodynamics, hence the funneling of luminous dwarf satellites into halos, on these scales requires relatively high numerical resolution far from the halos ($<30$ kpc at $> 1-2$ Mpc). This is typically out of the reach of most large volumes cosmological runs small-volume zoom likewise. In this regard, NewHorizon lies at a soft spot in the simulation landscape, which explains its ability to reproduce levels of lopsidedness rare in previous numerical studies.

\section{Discussion and Conclusion}
\label{sec:conclusion}

In this study, we investigated the prevalence and physical origin of satellite galaxy lopsidedness in 12 Milky Way-to-Cen A mass systems in the NewHorizon simulation. Reconstructing their surrounding gaseous cosmic web, we examined how the geometry and evolution of the cosmic environment around each system shape lopsided spatial distribution of satellites. Our main findings are:

\begin{itemize}

    \item Luminous satellite populations are more lopsided than expected if lopsidedness was simply constrained by sampling or the shape of their DM halo. These populations are also more lopsided than their DM sub-halo counterparts. Five of the twelve systems show broad angular overdensities, exceeding $2\sigma$ significance level, indicating that pronounced lopsidedness is not simply the result of a narrowly concentrated group of satellites.

    \item The orientation of lopsidedness is closely connected to the cosmic web. Planes of maximum lopsidedness align with nearby filaments, with a mean angle of $13.9^\circ$ across the sample ($p<0.001$). The corresponding association with the nearest massive neighboring galaxy is weaker, suggesting that their shared filamentary environment may help explain some observed companion-directed asymmetries.

    \item Filament connectivity and geometry regulate the strength of lopsidedness. Hemispheric asymmetry anti-correlates with filament connectivity ($\rho=-0.67$). Systems supplied by accretion channels distributed across multiple directions tend to be less lopsided, whereas channels concentrated within one hemisphere can produce pronounced asymmetries.

    \item The evolution of the filamentary network provides a mechanism for generating strong lopsidedness. In nodal systems, dominant accretion channels are concentrated on one side of the host, often in relation to a recent or on-going zipper (filament merger); in non-nodal systems, the relative position of the dominant (densest) filament shapes the asymmetry of the satellite distribution. This depends on how offset each system is from said main filament and whether it is embedded in a side stream of lower contrast, typically perpendicular to the main filament at large distance.

    \item M31's lopsidedness is compatible with the simulated luminous satellite population under our comparison. Depending on sample selection, M31 lies approximately $1$--$1.7\sigma$ from the inferred NewHorizon distribution. Although the limited number of simulated hosts and differences in satellite completeness prevent a definitive population-level assessment, this comparison highlights the importance of using resolved luminous satellite populations rather than simple dark matter-based models as null hypothesis when assessing the anomalousness of an observed system.

\end{itemize}

Together, these findings suggest that satellite lopsidedness is not solely a consequence of recent group infall or direct interactions with massive companions. Instead, it can emerge from the collective geometry and dynamical evolution of the filamentary network through which satellites are accreted. The spatial distribution of satellites thus retains information about the organization and assembly history of the cosmic web beyond the host halo. The same few filaments that funnel satellites also deliver cold gas and angular momentum to the central galaxy and set its spin orientation relative to the web \citep{Pichon11,Welker20}, so the perturbations the host disk experiences—minor mergers, flybys, and misaligned accretion—are likely to be as anisotropic and web-aligned as its satellites. Whether this imprint reaches the disk itself, for instance in its thickness, warps, or lopsidedness, is a natural next step that this census of zippers and twisters makes testable.

The spatial distribution of satellites thus retains information about the organization and assembly history of the cosmic web beyond the host halo. Low-connectivity systems receive most of their material through one or two persistent channels offset from the host (Section \ref{sec:zip}), the configuration in which inflowing gas should deliver angular momentum most coherently and the condition thought to build thin disks \citep{Pichon11}. The anti-correlation between lopsidedness and connectivity (Section \ref{sec:hemisphere}) therefore suggests that satellite lopsidedness may trace how coherently the host disk is fed, a link that can be tested against the disk’s thickness and rotational support.

An interesting question is whether the connections between filament connectivity, satellite anisotropy, and coherent accretion evolve across the characteristic golden mass, $M_{\rm halo}\sim10^{12},M_\odot$, associated with a transition in the dominant physical processes regulating galaxy growth \citep{Dekel2019, 2006MNRAS.368....2D}. In \cite{Madhani2026Zippers}, we identified a sharp transition in filament connectivity and satellite-plane alignment across this mass scale, which we interpreted in the context of the broader physical transitions occurring in this regime. Our present sample spans this characteristic mass, where the interplay between cold filamentary accretion, virial shock heating, and feedback becomes particularly important. While the limited sample size precludes us from establishing a corresponding mass-dependent transition in satellite lopsidedness, our results suggest that the geometry and evolution of the surrounding filamentary network may also play a role in shaping satellite anisotropy across this regime.  Extending this analysis to a larger population of halos would allow us to investigate whether changes in the efficiency and coherence of filamentary accretion across the golden mass are accompanied by systematic changes in satellite anisotropy.

Our results also highlight the importance of resolving the gaseous environment on scales extending well beyond the virial radius. Accurately capturing satellite anisotropies requires simulations that resolve not only low-mass galaxies and their host halos, but also the intermediate-scale filaments and streams that regulate their accretion. This picture is consistent with the Lagrangian theoretical analysis of \citet{shim2026filamentsmatterrelativescalings}, which predicts that anisotropic infall into halos is primarily driven by tidal contributions from a characteristic, mass-dependent range of intermediate scales, extending beyond the nearest neighbors but becoming relatively insensitive to  more distant structures ($\gtrsim 10$ Mpc). Our findings provide a complementary hydrodynamic perspective on this preferred range of environmental scales that influence anisotropic infall. To date, such simulated cosmic volumes are rare and still relatively small. However, together with planar satellite configurations, lopsidedness provides a complementary probe of how the evolving cosmic web shapes the assembly of galaxies and their satellite populations. The design of larger volumes, encompassing denser regions of the Universe is therefore a key target for the next decade, as dwarf galaxies become of prime focus of modern telescopes like LSST/Rubin.

While the spatial alignments and evolutionary histories identified here provide strong evidence for a filamentary origin, a more direct reconstruction of individual satellite accretion histories will be necessary to quantify the relative contributions of filament mergers, coherent accretion, group infall, and subsequent orbital evolution. Such an analysis will also need to establish how long the resulting asymmetries persist and under what conditions they are erased or reinforced.

\section{Acknowledgments}
We thank Mia de Los Reyes and Joanna Woo for valuable discussions.
The core scientific team is comprised of Janvi Madhani, Charlotte Welker and Sneha Nair. JM led the analysis, implemented the pipeline in Python, produced most analysis plots, wrote parts of the draft. CW designed the study and the overall  pipeline, wrote parts of the draft and supervised the core team. SN developed and optimized a Python package to analyze gas properties around filaments and produced stream profiles. All authors commented on the final draft. JM and CW acknowledge the support of the LSST-DA Catalyst Fellowship funded by the LSST Discovery Alliance and the Templeton Foundation and of the LSST-DA Early Science award funded by the LSST Discovery Alliance and the Simons Foundation. SN and CW acknowledge the support of NSF LEAPS-MPS Award AST-2316862. Main computational resources were provided by CP and YD through access to Infinity HPC cluster at IAP managed by Stephane Rouberol and to the NewHorizon and Horizon-AGN simulations. CP and KK acknowledge the support of French ANR grant GALBAR ANR-25-CE31-468. This project has also received financial support from the CNRS through the MITI interdisciplinary programs. No generative AI tool was used in the design, data preparation, data analysis, figure preparation, interpretation or writing of this manuscript. One co-author who commented on the final draft reported using an AI tool for detection of typos and minor inaccuracies as part of his comments.

\appendix

\section{Variations in Maximizing Plane Orientations}
\label{appendix:degeneracy}

\begin{figure*}
\centering
	\includegraphics[width=0.9\textwidth]{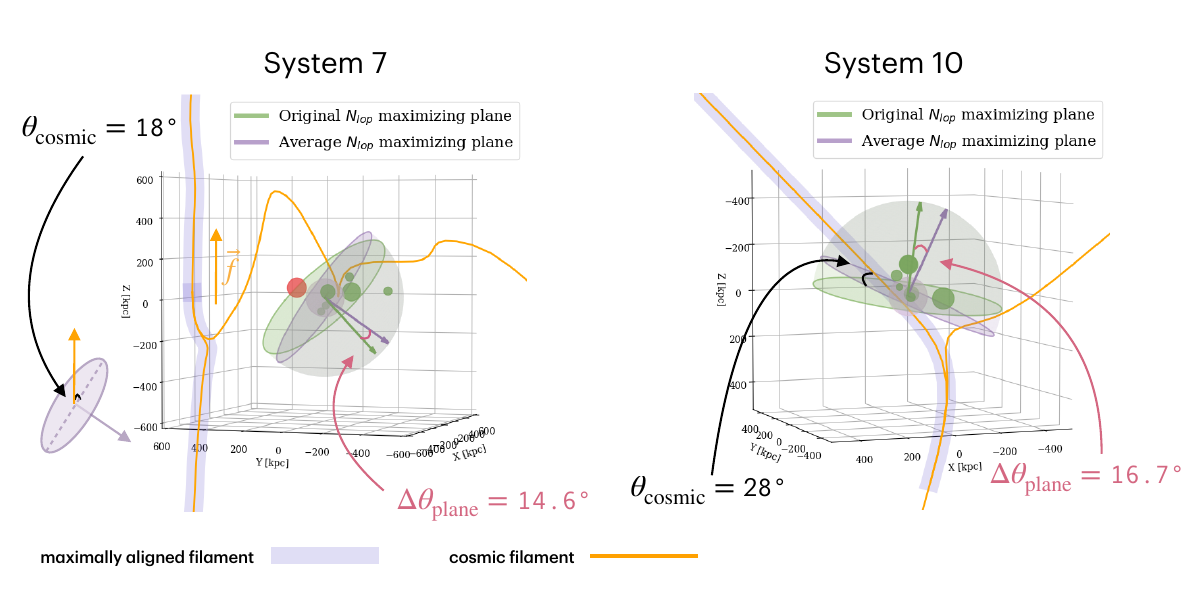}
    \caption{ Distributions of satellites and nearby cosmic filaments for the two least populated NewHorizon systems ($N_{\rm sat}<10$). Green surfaces indicate the original maximizing planes selected from 10,000 randomly sampled trial orientations, while purple surfaces indicate the average alternative orientation yielding the same maximum hemispheric satellite count ($N_{\rm lop}$). The angular separations between the maximizing planes are $\Delta\theta_{\rm plane}=14.6^\circ$ and $16.7^\circ$ for Systems 7 and 10, respectively. Green and red circles represent satellites within and outside the populated hemisphere of the original plane, while the central galaxy is shown in light purple. Orange curves trace nearby cosmic filaments, with the most closely aligned filament with respect to the average maximizing plane highlighted in light purple. Black arrows indicate the filament-alignment angles, $\theta_{\rm cosmic}$. Pink arrows indicate the angular separation, $\Delta \theta_{\rm plane}$, between the original maximizing plane and the average one. }
    \label{fig:syst7_10_degen}
\end{figure*}

As discussed in Section~\ref{sec:plane_max_asym}, the orientation of the plane of maximum asymmetry is not necessarily unique for a finite satellite population. Across our sample, the maximum angular separation between orientations yielding the same $N_{\rm lop}$ remains below our adopted angular resolution of $\Delta\theta=10^\circ$ for all but Systems 7 and 10, the two least populated systems ($N_{\rm sat}<10$) which have maximum angular separations between $\sim15-20^\circ$. Their larger variations arise from the limited number of satellites constraining the plane orientation, as illustrated in Fig.~\ref{fig:syst7_10_degen}. Adopting a representative orientation derived from the ensemble of maximizing planes would strengthen their apparent alignment with the nearby cosmic filaments, reducing $\theta_{\rm cosmic}$ from $22.49^\circ$ to $17.94^\circ$ for System 7 and from $42.75^\circ$ to $27.74^\circ$ for System 10. Nevertheless, we retain the first maximizing orientation identified among our randomly sampled trial normals for all systems, ensuring a consistent selection procedure that is independent of the surrounding cosmic-web geometry and does not preferentially enhance the measured filament alignments.

\section{Is Andromeda anomalously lopsided compared to NewHorizon?}
\label{appendix:m31}

One question that arises is whether, in light of NewHorizon results, Andromeda's satellites can still be considered an anomalously lopsided system, at odds with $\Lambda$CDM cosmology. 
An issue is that no NewHorizon system has exactly 37 satellites as Andromeda and we have shown throughout the study that under-sampling a rich system is not equivalent to a system with organically $N_{\rm sat}=37$. However, the satellite masses probed in \cite{Kanehisa_2025} are also likely slightly more biased towards smaller masses than NewHorizon, which misses satellites with $M_{*}<10^{5.5}M_{\odot}$. By comparison, the smallest mass detected for an M31 satellite is $\sim 2-3\times 10^4\,\rm M_{\odot}$ (corresponding to only 2 stellar particles in NewHorizon), although surveys remain starkly incomplete in this regime. M31 may therefore be more comparable to medium richness systems in NewHorizon.

\begin{figure*}
\centering
	\includegraphics[width=0.9\textwidth]{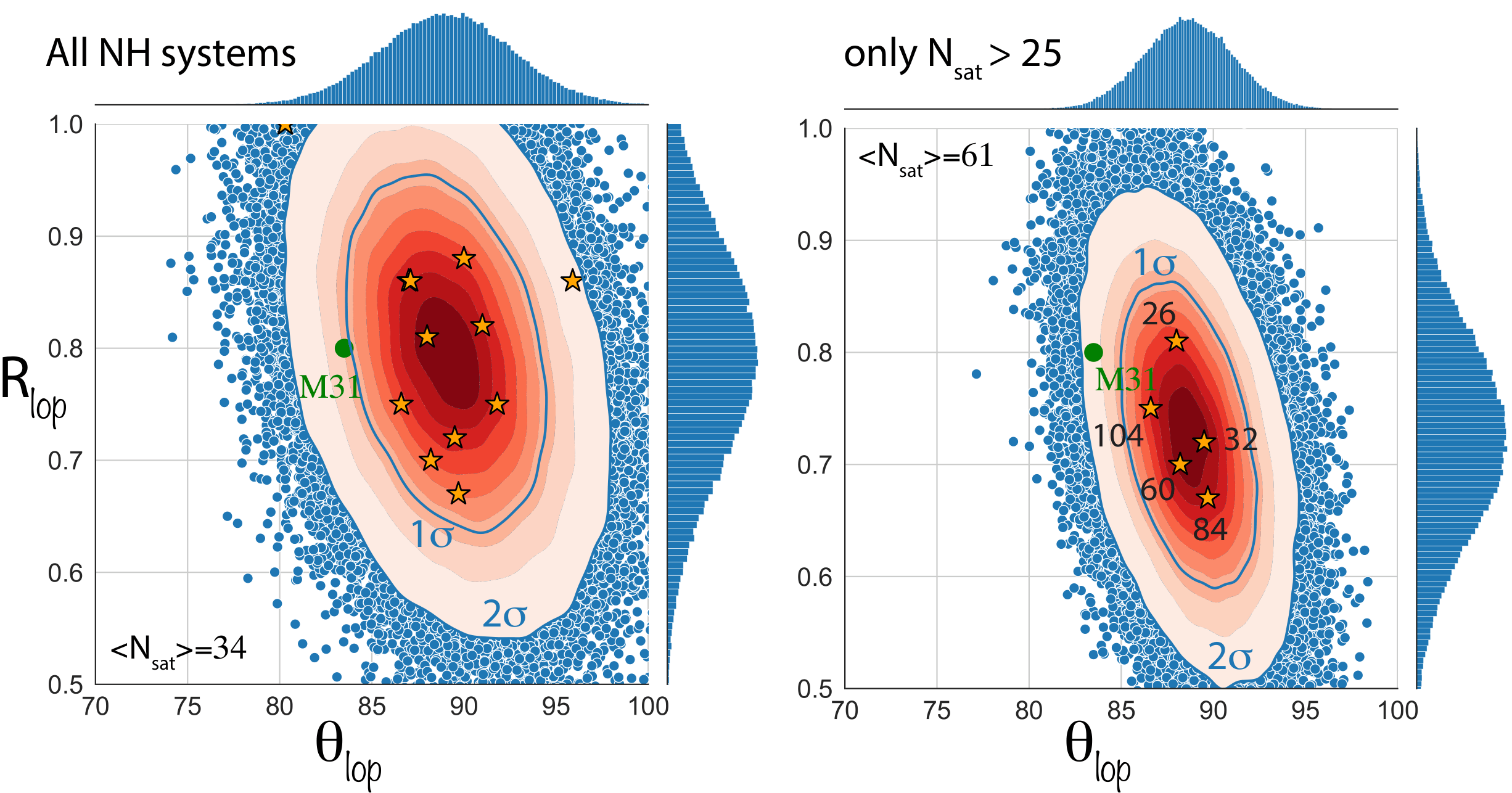}
    \caption{ Mean posterior joint distribution for $R_{\rm lop}$ and $\theta_{\rm lop}$ from priors emulating the halo-tracing hypothesis in NewHorizon. The posterior is updated with the measured parameters of: either all the 11 NewHorizon local massive galaxy analogs ({\it Left}), or restricted to systems with $N_{\rm sat}>25$ ({\it Right}). 1- and 2-$\sigma$ contours are shown in blue. The NewHorizon systems appear as yellow stars and M31 as a green dot. M31 lopsidedness is fully compatible with NewHorizon.}
    \label{fig:Andromeda}
\end{figure*}

To propose a tentative answer to this question, we perform a population-level Bayesian inference of what the joint distribution of $R_{\rm lop}$ and $\theta_{\rm lop}$ should be in the population of MW-mass and above galaxies of the nearby Universe, assuming that the NewHorizon sample is representative of this population. The Bayesian framework allows us to take the halo-tracing hypothesis of Fig.~\ref{fig:rlop_vs_thetalop} as a prior, to avoid overestimating the predictive power of NewHorizon. ($R_{\rm lop}$, $\theta_{\rm lop}$) data-points from the NewHorizon samples are then used to update likelihood and posterior distributions of $R_{\rm lop}$ and $\theta_{\rm lop}$. The methodology is similar to the one we described in our previous study \citep{Madhani2026Zippers}. We refer the reader to it for more details. The only differences is that the parametric form of the joint ($R_{\rm lop}$, $\theta_{\rm lop}$) distribution is simply assumed to be a bi-variate gaussian with means ($\mu_1,\mu_2$), standard deviations ($\sigma_1,\sigma_2$) and correlation coefficient $\rho$, in line with their observed distributions in NewHorizon. This form is assumed for both the prior and the posterior distributions, with prior parameters fitted to the null-hypothesis of  Fig.~\ref{fig:rlop_vs_thetalop}, left panel, and posterior values derived by the Bayesian inference scheme.

Fig.~\ref{fig:Andromeda} shows the result of the inference using all NewHorizon systems on the left panel (leading to an average $N_{\rm sat}=34$) and restricted to systems with more than 25 satellites (leading to an average $N_{\rm sat}=61$). Note that the differences in satellite mass sensitivity between NewHorizon and M31 do not allow to conclude easily about which case is the most comparable to the M31 halo, we therefore produce both figures. 

Irrespective of NewHorizon selection, the same conclusion is reached: the lopsidedness of M31 is not particularly remarkable compared to similar NewHorizon systems with currently available information. It is barely past the $1\sigma$ contour of the NewHorizon posterior distribution in the first case and only reaches $1.7\sigma$ in the second case. Therefore we conclude that the significance of M31 lopsidedness has been overestimated by previous comparisons to simulations as these were under-resolved and/or were assessed through resampling of richer systems to M31 richness, a procedure that erases correlations between richness and environment as we already demonstrated in \cite{Madhani2026Zippers}.

\section{Effect of sampling bias}
\label{appendix:sampling-bias}
\begin{figure}
\centering
	\includegraphics[width=1\columnwidth]{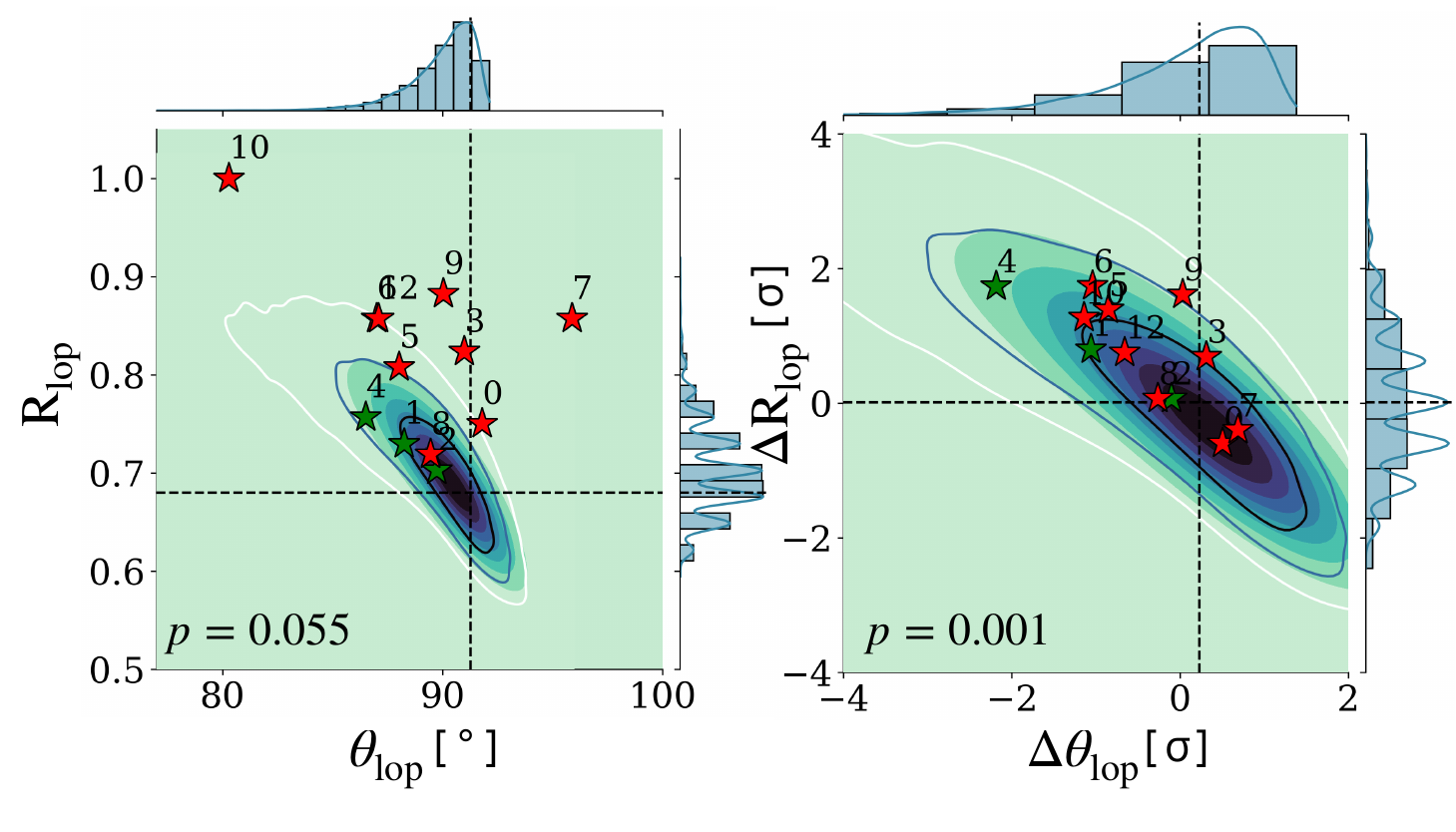}
    \caption{{\it Left:} Same as Fig.~\ref{fig:rlop_vs_thetalop} but after resampling systems to $N_{\rm sat}=37$ and taking the means. Systems with strictly less than 37 satellites are indicated as red stars for reference but not included in the comparison. {\it Right:} Same analysis for $\Delta R_{\rm lop}$ and $\Delta \theta_{\rm lop}$}
    \label{fig:rlop_thetalop}
\end{figure}

To better constrain the effect of satellite sampling on the general trends observed and to allow for a better comparison with M31, we repeat the analysis but sub-sampling every sufficiently rich NewHorizon system to $N_{\rm sat}=37$. Specifically, each sufficiently rich system is randomly sub-sampled 1000 times and resulting means are presented on Fig.~\ref{fig:rlop_thetalop}, left panel, while relative deviations to individual null-hypothesis are presented on the right panel. Rich resampled systems are presented in green. Non-resampled systems with $N_{\rm sat}<37$ are still displayed in red for reference and easy comparison to Fig.~\ref{fig:rlop_vs_thetalop} but not included in the null-hypothesis.

While averaging each system over multiple re-samplings unsurprisingly reduces mean $R_{\rm lop}$ values, results are still qualitatively the same for these rich systems most comparable to Andromeda: their lopsidedness is inconsistent with the halo-tracing hypothesis and systematically more extreme, even after resampling and averaging.

While we have too few rich systems to obtain a large number of high deviation systems in our sample,  this indicates once again that using isotropic or even halo-tracing distributions as null hypothesis will systematically over-estimate how unusual and anomalous an observed lopsided system is. A better understanding of how lopsidedness arises in halos, in particular at the baryonic level, is needed to design more relevant null-hypothesis.

\section{NewHorizon Systems: Lopsidedness and Alignment with Cosmic Filaments}
\label{appendix:alignments}

\subsection{Hemispheric lopsidedness}

\begin{figure*}
\centering
	\includegraphics[width=\textwidth]{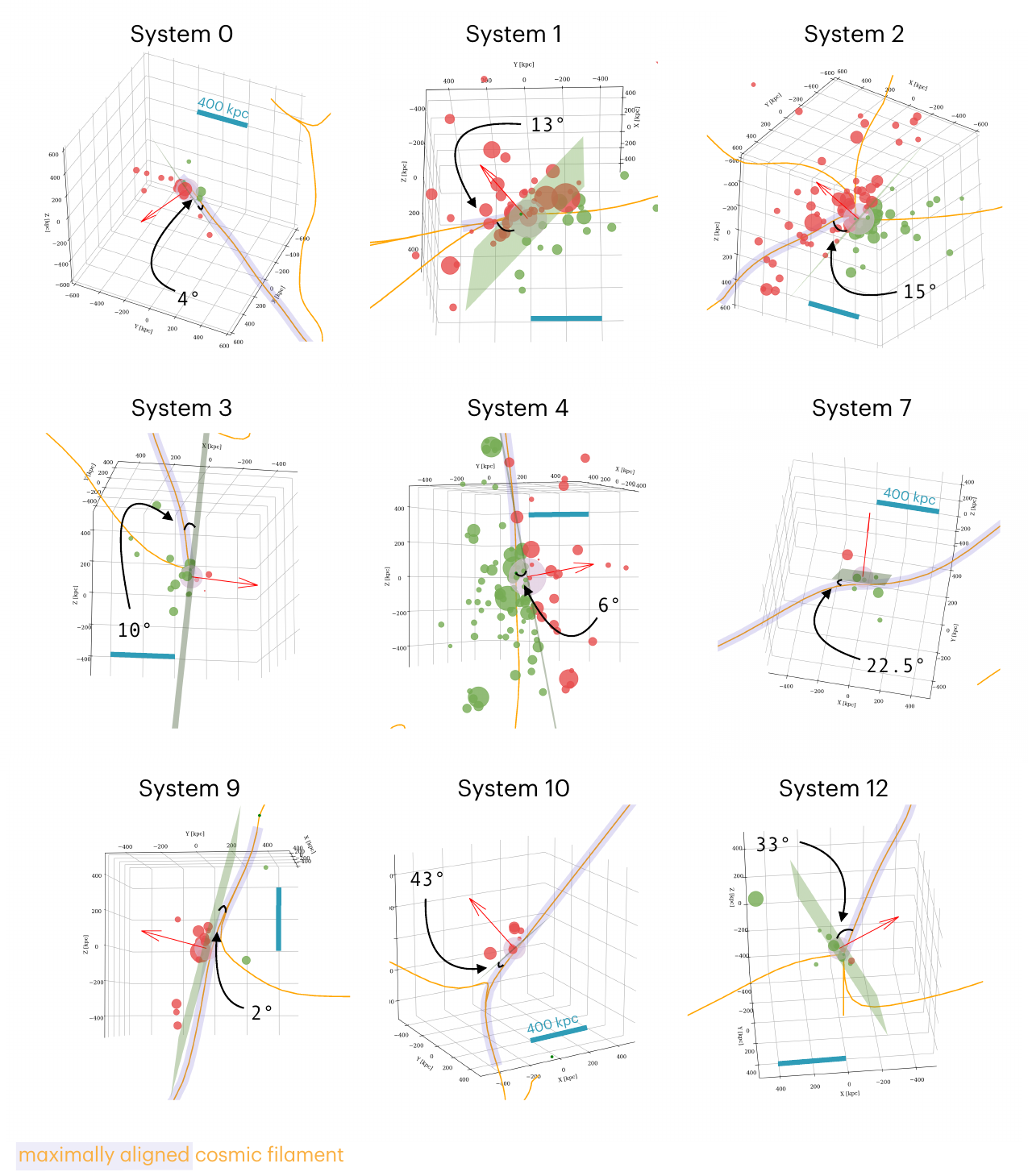}
    \caption{Same as Figure \ref{fig:maxalign_fil} but for all 9 systems in our sample not presented in the main text. The plane of maximum asymmetry is shown in light green, $\vec{n}_{\rm lop}$ as a red arrow and satellites as red/green spheres, scaled by mass. The central galaxy appears as light purple. The angle between the plane and most aligned cosmic filament, $\theta_{\rm cosmic}$, is marked in black. Filaments identified by \texttt{DisPerSE} are displayed in orange, with the maximally aligned one shaded in purple grey. }
    \label{fig:allfil}
\end{figure*}

Fig.~\ref{fig:allfil} displays 3D maps of the 9 systems of satellites not otherwise displayed in Fig.~\ref{fig:maxalign_fil}, with their plane of maximum asymmetry highlighted in light green and cosmic filaments in orange. The most aligned cosmic filament is shaded in grey.
As discussed in the main text, strong alignment of the plane of maximum asymmetry with a nearby cosmic filament is ubiquitous across the sample.

\subsection{Focal lopsidedness}

\begin{figure*}
	\includegraphics[width=\linewidth]{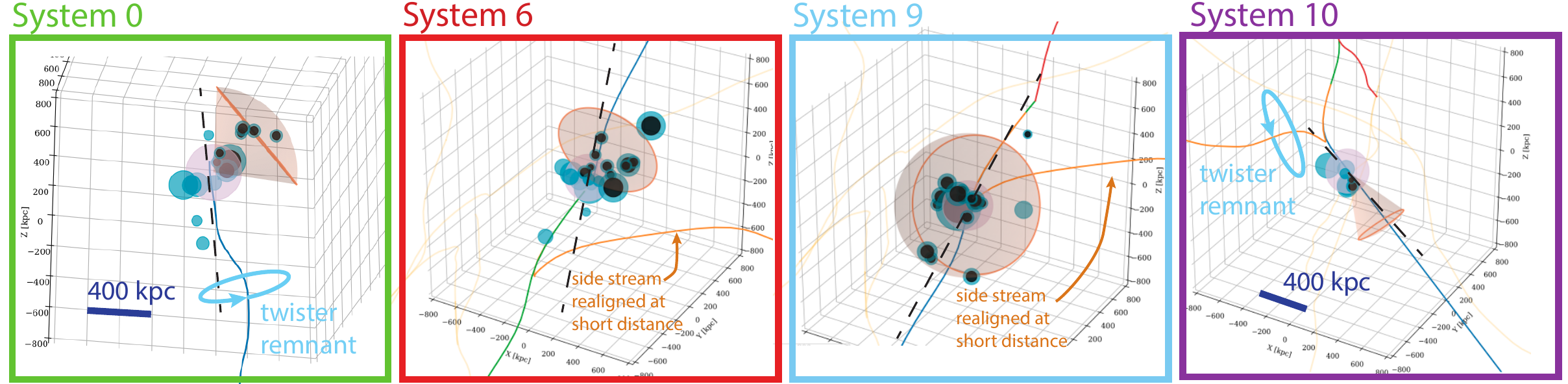}
    \caption{3D maps of cosmic filaments, satellites and cones of maximal lopsidedness around non-nodal systems in NewHorizon. Zippers and twisters (filament mergers with and without orbital momentum) appear as blue arrows. The intersection of the plane of maximum asymmetry with the projection plane centered on the central is overlaid as dashed line for reference.}
    \label{fig:non-nodal}
\end{figure*}

Fig.~\ref{fig:non-nodal} displays 3D maps of the non-nodal systems of satellites not otherwise displayed in the main text, with their cone of maximal lopsidedness highlighted in light orange and cosmic filaments as solid colored lines. 

Narrow cones of maximal lopsidedness are directly aligned with the dominant filament in 2/3 cases. Note how System 9 displays a broad cone of maximal lopsidedness, which includes the long, straight filament that the system lives on the midst of. Only System 0 displays a cone of maximal lopsidedness misaligned with its visible filaments. Further analysis reveals it is aligned with recent group accretion from a side stream just below the persistence cut-off of our filament detection.







\bibliography{bibliography}{}

\begin{thebibliography}{}
\expandafter\ifx\csname natexlab\endcsname\relax\def\natexlab#1{#1}\fi
\providecommand{\url}[1]{\href{#1}{#1}}
\providecommand{\dodoi}[1]{doi:~\href{http://doi.org/#1}{\nolinkurl{#1}}}
\providecommand{\doeprint}[1]{\href{http://ascl.net/#1}{\nolinkurl{http://ascl.net/#1}}}
\providecommand{\doarXiv}[1]{\href{https://arxiv.org/abs/#1}{\nolinkurl{https://arxiv.org/abs/#1}}}

\bibitem[{S.~H. {Ahmed} {et~al.}(2017){Ahmed}, {Brooks}, \& {Christensen}}]{2017MNRAS.466.3119A}
{Ahmed}, S.~H., {Brooks}, A.~M., \& {Christensen}, C.~R. 2017, \bibinfo{title}{{The role of baryons in creating statistically significant planes of satellites around Milky Way-mass galaxies},} \mnras, 466, 3119, \dodoi{10.1093/mnras/stw3271}

\bibitem[{D. Aubert {et~al.}(2004)Aubert, Pichon, \& Colombi}]{Aubert_2004}
Aubert, D., Pichon, C., \& Colombi, S. 2004, \bibinfo{title}{The origin and implications of dark matter anisotropic cosmic infall on $\approx$$\less$i$\greater$L$\less$/i$\greater$$\less$sub$\greater${\ding{72} }$\less$/sub$\greater$haloes,} Monthly Notices of the Royal Astronomical Society, 352, 376, \dodoi{10.1111/j.1365-2966.2004.07883.x}

\bibitem[{S. Barsanti {et~al.}(2022)Barsanti, Colless, Welker, Oh, Casura, Bryant, Croom, D'Eugenio, Lawrence, Richards, \& van~de Sande}]{Barsanti2022}
Barsanti, S., Colless, M., Welker, C., {et~al.} 2022, \bibinfo{title}{The SAMI Galaxy Survey: flipping of the spin-filament alignment correlates most strongly with growth of the bulge,} Monthly Notices of the Royal Astronomical Society, 516, 3569, \dodoi{10.1093/mnras/stac2405}

\bibitem[{A. Bowden {et~al.}(2014)Bowden, Evans, \& Belokurov}]{Bowden_2014}
Bowden, A., Evans, N.~W., \& Belokurov, V. 2014, \bibinfo{title}{ON ASYMMETRIC DISTRIBUTIONS OF SATELLITE GALAXIES,} The Astrophysical Journal, 793, L42, \dodoi{10.1088/2041-8205/793/2/l42}

\bibitem[{T.~G. Brainerd \& A. Samuels(2020)Brainerd \& Samuels}]{Brainerd_2020}
Brainerd, T.~G., \& Samuels, A. 2020, \bibinfo{title}{Lopsided Satellite Distributions around Isolated Host Galaxies,} The Astrophysical Journal Letters, 898, L15, \dodoi{10.3847/2041-8213/aba194}

\bibitem[{A. {Dekel} \& Y. {Birnboim}(2006){Dekel} \& {Birnboim}}]{2006MNRAS.368....2D}
{Dekel}, A., \& {Birnboim}, Y. 2006, \bibinfo{title}{{Galaxy bimodality due to cold flows and shock heating},} \mnras, 368, 2, \dodoi{10.1111/j.1365-2966.2006.10145.x}

\bibitem[{A. {Dekel} {et~al.}(2019){Dekel}, {Sarkar}, {Jiang}, {Bournaud}, {Krumholz}, {Ceverino}, \& {Primack}}]{Dekel2019}
{Dekel}, A., {Sarkar}, K.~C., {Jiang}, F., {et~al.} 2019, \bibinfo{title}{{The global star formation law by supernova feedback},} MNRAS, 488, 4753, \dodoi{10.1093/mnras/stz1919}

\bibitem[{Y. {Dubois} {et~al.}(2014){Dubois}, {Pichon}, {Welker}, {Le Borgne}, {Devriendt}, {Laigle}, {Codis}, {Pogosyan}, {Arnouts}, {Benabed}, {Bertin}, {Blaizot}, {Bouchet}, {Cardoso}, {Colombi}, {de Lapparent}, {Desjacques}, {Gavazzi}, {Kassin}, {Kimm}, {McCracken}, {Milliard}, {Peirani}, {Prunet}, {Rouberol}, {Silk}, {Slyz}, {Sousbie}, {Teyssier}, {Tresse}, {Treyer}, {Vibert}, \& {Volonteri}}]{2014MNRAS.444.1453D}
{Dubois}, Y., {Pichon}, C., {Welker}, C., {et~al.} 2014, \bibinfo{title}{{Dancing in the dark: galactic properties trace spin swings along the cosmic web},} \mnras, 444, 1453, \dodoi{10.1093/mnras/stu1227}

\bibitem[{Y. {Dubois} {et~al.}(2021){Dubois}, {Beckmann}, {Bournaud}, {Choi}, {Devriendt}, {Jackson}, {Kaviraj}, {Kimm}, {Kraljic}, {Laigle}, {Martin}, {Park}, {Peirani}, {Pichon}, {Volonteri}, \& {Yi}}]{2021A&A...651A.109D}
{Dubois}, Y., {Beckmann}, R., {Bournaud}, F., {et~al.} 2021, \bibinfo{title}{{Introducing the NEWHORIZON simulation: Galaxy properties with resolved internal dynamics across cosmic time},} \aap, 651, A109, \dodoi{10.1051/0004-6361/202039429}

\bibitem[{D. Gal{\'a}rraga-Espinosa {et~al.}(2022)Gal{\'a}rraga-Espinosa, Langer, \& Aghanim}]{GalarragaEspinosa2022}
Gal{\'a}rraga-Espinosa, D., Langer, M., \& Aghanim, N. 2022, \bibinfo{title}{Relative distribution of dark matter, gas, and stars around cosmic filaments in the {IllustrisTNG} simulation,} Astronomy \& Astrophysics, 661, A115, \dodoi{10.1051/0004-6361/202141974}

\bibitem[{C.~C. Gong {et~al.}(2019)Gong, Libeskind, Tempel, Guo, Gottl{\"o}ber, Yepes, Wang, Sorce, \& Pawlowski}]{Gong_2019}
Gong, C.~C., Libeskind, N.~I., Tempel, E., {et~al.} 2019, \bibinfo{title}{The origin of lopsided satellite galaxy distribution in galaxy pairs,} Monthly Notices of the Royal Astronomical Society, 488, 3100, \dodoi{10.1093/mnras/stz1917}

\bibitem[{Y. {Guo} {et~al.}(2025){Guo}, {Ma}, \& {Li}}]{Guo2025}
{Guo}, Y., {Ma}, Q., \& {Li}, C. 2025, \bibinfo{title}{{Lopsided and Bulging Distribution of Satellites Around Paired Halos. I. Observational Measurements and Comparison with Halo-based Models},} \apj, 992, 157, \dodoi{10.3847/1538-4357/ae00c3}

\bibitem[{N. Heesters {et~al.}(2024)Heesters, Jerjen, Müller, Pawlowski, \& Kanehisa}]{Heesters_2024}
Heesters, N., Jerjen, H., Müller, O., Pawlowski, M.~S., \& Kanehisa, K.~J. 2024, \bibinfo{title}{Lopsided distribution of MATLAS and ELVES dwarf satellite systems around isolated host galaxies,} Astronomy and Astrophysics, 690, A110, \dodoi{10.1051/0004-6361/202450110}

\bibitem[{H. {Jhee} {et~al.}(2026){Jhee}, {Song}, {Laigle}, {Pichon}, {Cadiou}, \& {Choi}}]{Jhee26}
{Jhee}, H., {Song}, H., {Laigle}, C., {et~al.} 2026, \bibinfo{title}{{The impact of evolving cosmic filaments on mass and spin evolution of dark matter halos},} arXiv e-prints, arXiv:2606.19443, \dodoi{10.48550/arXiv.2606.19443}

\bibitem[{M.~P. {J{\'u}lio} {et~al.}(2024){J{\'u}lio}, {Pawlowski}, {Tony Sohn}, {Taibi}, {van der Marel}, \& {McGaugh}}]{2024A&A...687A.212J}
{J{\'u}lio}, M.~P., {Pawlowski}, M.~S., {Tony Sohn}, S., {et~al.} 2024, \bibinfo{title}{{Satellite group infall into the Milky Way: Exploring the Crater-Leo case with new HST proper motions},} \aap, 687, A212, \dodoi{10.1051/0004-6361/202449985}

\bibitem[{K.~J. Kanehisa {et~al.}(2025)Kanehisa, Pawlowski, \& Libeskind}]{Kanehisa_2025}
Kanehisa, K.~J., Pawlowski, M.~S., \& Libeskind, N. 2025, \bibinfo{title}{Andromeda’s asymmetric satellite system as a challenge to cold dark matter cosmology,} Nature Astronomy, 9, 692–701, \dodoi{10.1038/s41550-025-02480-3}

\bibitem[{T. {Kimm} {et~al.}(2015){Kimm}, {Cen}, {Devriendt}, {Dubois}, \& {Slyz}}]{2015MNRAS.451.2900K}
{Kimm}, T., {Cen}, R., {Devriendt}, J., {Dubois}, Y., \& {Slyz}, A. 2015, \bibinfo{title}{{Towards simulating star formation in turbulent high-z galaxies with mechanical supernova feedback},} \mnras, 451, 2900, \dodoi{10.1093/mnras/stv1211}

\bibitem[{T. {Kimm} {et~al.}(2011){Kimm}, {Devriendt}, {Slyz}, {Pichon}, {Kassin}, \& {Dubois}}]{2011arXiv1106.0538K}
{Kimm}, T., {Devriendt}, J., {Slyz}, A., {et~al.} 2011, \bibinfo{title}{{The angular momentum of baryons and dark matter halos revisited},} arXiv e-prints, arXiv:1106.0538, \dodoi{10.48550/arXiv.1106.0538}

\bibitem[{E. {Komatsu} {et~al.}(2011){Komatsu}, {Smith}, {Dunkley}, {Bennett}, {Gold}, {Hinshaw}, {Jarosik}, {Larson}, {Nolta}, {Page}, {Spergel}, {Halpern}, {Hill}, {Kogut}, {Limon}, {Meyer}, {Odegard}, {Tucker}, {Weiland}, {Wollack}, \& {Wright}}]{2011ApJS..192...18K}
{Komatsu}, E., {Smith}, K.~M., {Dunkley}, J., {et~al.} 2011, \bibinfo{title}{{Seven-year Wilkinson Microwave Anisotropy Probe (WMAP) Observations: Cosmological Interpretation},} \apjs, 192, 18, \dodoi{10.1088/0067-0049/192/2/18}

\bibitem[{K. {Kraljic} {et~al.}(2021){Kraljic}, {Duckworth}, {Tojeiro}, {Alam}, {Bizyaev}, {Weijmans}, {Boardman}, \& {Lane}}]{Kraljic21}
{Kraljic}, K., {Duckworth}, C., {Tojeiro}, R., {et~al.} 2021, \bibinfo{title}{{SDSS-IV MaNGA: 3D spin alignment of spiral and S0 galaxies},} \mnras, 504, 4626, \dodoi{10.1093/mnras/stab1109}

\bibitem[{Y.-S. {Li} \& A. {Helmi}(2008){Li} \& {Helmi}}]{2008MNRAS.385.1365L}
{Li}, Y.-S., \& {Helmi}, A. 2008, \bibinfo{title}{{Infall of substructures on to a Milky Way-like dark halo},} \mnras, 385, 1365, \dodoi{10.1111/j.1365-2966.2008.12854.x}

\bibitem[{N.~I. Libeskind {et~al.}(2016)Libeskind, Guo, Tempel, \& Ibata}]{Libeskind_2016}
Libeskind, N.~I., Guo, Q., Tempel, E., \& Ibata, R. 2016, \bibinfo{title}{The lopsided distribution of satellite galaxies,} \doarXiv{1606.01516}

\bibitem[{N.~I. {Libeskind} {et~al.}(2014){Libeskind}, {Knebe}, {Hoffman}, \& {Gottl{\"o}ber}}]{2014MNRAS.443.1274L}
{Libeskind}, N.~I., {Knebe}, A., {Hoffman}, Y., \& {Gottl{\"o}ber}, S. 2014, \bibinfo{title}{{The universal nature of subhalo accretion},} \mnras, 443, 1274, \dodoi{10.1093/mnras/stu1216}

\bibitem[{Y. Liu {et~al.}(2024)Liu, Wang, Guo, Springel, Bose, Pakmor, \& Hernquist}]{Liu_2024}
Liu, Y., Wang, P., Guo, H., {et~al.} 2024, \bibinfo{title}{The origin of lopsided satellite galaxy distribution around isolated systems in MillenniumTNG,} Monthly Notices of the Royal Astronomical Society, 529, 1405, \dodoi{10.1093/mnras/stae625}

\bibitem[{Y.~S. {Lu} {et~al.}(2024){Lu}, {Mandelker}, {Oh}, {Dekel}, {van den Bosch}, {Springel}, {Nagai}, \& {van de Voort}}]{Lu2024}
{Lu}, Y.~S., {Mandelker}, N., {Oh}, S.~P., {et~al.} 2024, \bibinfo{title}{{The structure and dynamics of massive high-z cosmic-web filaments: three radial zones in filament cross-sections},} \mnras, 527, 11256, \dodoi{10.1093/mnras/stad3779}

\bibitem[{Q. {Ma} {et~al.}(2025){Ma}, {Li}, \& {Guo}}]{2025ApJ...992...41M}
{Ma}, Q., {Li}, C., \& {Guo}, Y. 2025, \bibinfo{title}{{Lopsided and Bulging Distribution of Satellites around Paired Halos. II. 3D Analysis and Dependence on Projection and Selection Effects},} \apj, 992, 41, \dodoi{10.3847/1538-4357/adfede}

\bibitem[{J.~P. Madhani {et~al.}(2026)Madhani, Welker, Nair, Gallego, Feliciano, Pichon, Olsen, Dubois, Kaviraj, \& Kraljic}]{Madhani2026Zippers}
Madhani, J.~P., Welker, C., Nair, S., {et~al.} 2026, \bibinfo{title}{Zippers and Twisters: Planes of Satellite Galaxies Emerge from Whirling and Shocking Gas Streams in the Cosmic Web,} The Astrophysical Journal, 1007, 86, \dodoi{10.3847/1538-4357/ae81b4}

\bibitem[{M.~S. Pawlowski {et~al.}(2017)Pawlowski, Ibata, \& Bullock}]{Pawlowski_2017}
Pawlowski, M.~S., Ibata, R.~A., \& Bullock, J.~S. 2017, \bibinfo{title}{The Lopsidedness of Satellite Galaxy Systems in {$\Lambda$CDM} Simulations,} The Astrophysical Journal, 850, 132, \dodoi{10.3847/1538-4357/aa9435}

\bibitem[{C. {Pichon} {et~al.}(2011){Pichon}, {Pogosyan}, {Kimm}, {Slyz}, {Devriendt}, \& {Dubois}}]{Pichon11}
{Pichon}, C., {Pogosyan}, D., {Kimm}, T., {et~al.} 2011, \bibinfo{title}{{Rigging dark haloes: why is hierarchical galaxy formation consistent with the inside-out build-up of thin discs?},} \mnras, 418, 2493, \dodoi{10.1111/j.1365-2966.2011.19640.x}

\bibitem[{J. Shim {et~al.}(2026)Shim, Pogosyan, Jeon, \& Pichon}]{shim2026filamentsmatterrelativescalings}
Shim, J., Pogosyan, D., Jeon, M., \& Pichon, C. 2026, Which filaments matter: the relative scalings of anisotropic infall, \doarXiv{2604.03093}

\bibitem[{H. {Song} {et~al.}(2021){Song}, {Laigle}, {Hwang}, {Devriendt}, {Dubois}, {Kraljic}, {Pichon}, {Slyz}, \& {Smith}}]{Song21}
{Song}, H., {Laigle}, C., {Hwang}, H.~S., {et~al.} 2021, \bibinfo{title}{{Beyond halo mass: quenching galaxy mass assembly at the edge of filaments},} \mnras, 501, 4635, \dodoi{10.1093/mnras/staa3981}

\bibitem[{T. {Sousbie} {et~al.}(2011){Sousbie}, {Pichon}, \& {Kawahara}}]{2011MNRAS.414..384S}
{Sousbie}, T., {Pichon}, C., \& {Kawahara}, H. 2011, \bibinfo{title}{{The persistent cosmic web and its filamentary structure - II. Illustrations},} \mnras, 414, 384, \dodoi{10.1111/j.1365-2966.2011.18395.x}

\bibitem[{E. Tempel \& A. Tamm(2015)Tempel \& Tamm}]{Tempel_2015}
Tempel, E., \& Tamm, A. 2015, \bibinfo{title}{Galaxy pairs align with Galactic filaments,} Astronomy \& Astrophysics, 576, L5, \dodoi{10.1051/0004-6361/201525827}

\bibitem[{R. {Teyssier}(2002){Teyssier}}]{2002A&A...385..337T}
{Teyssier}, R. 2002, \bibinfo{title}{{Cosmological hydrodynamics with adaptive mesh refinement. A new high resolution code called RAMSES},} \aap, 385, 337, \dodoi{10.1051/0004-6361:20011817}

\bibitem[{D. {Tweed} {et~al.}(2009){Tweed}, {Devriendt}, {Blaizot}, {Colombi}, \& {Slyz}}]{2009A&A...506..647T}
{Tweed}, D., {Devriendt}, J., {Blaizot}, J., {Colombi}, S., \& {Slyz}, A. 2009, \bibinfo{title}{{Building merger trees from cosmological N-body simulations. Towards improving galaxy formation models using subhaloes},} \aap, 506, 647, \dodoi{10.1051/0004-6361/200911787}

\bibitem[{P. Wang {et~al.}(2020)Wang, Libeskind, Tempel, Pawlowski, Kang, \& Guo}]{wang2020alignmentsatellitesystemscosmic}
Wang, P., Libeskind, N.~I., Tempel, E., {et~al.} 2020, The alignment of satellite systems with cosmic filaments in the SDSS DR12, \doarXiv{2007.08345}

\bibitem[{C. {Welker} {et~al.}(2020){Welker}, {Bland-Hawthorn}, {van de Sande}, {Lagos}, {Elahi}, {Obreschkow}, {Bryant}, {Pichon}, {Cortese}, {Richards}, {Croom}, {Goodwin}, {Lawrence}, {Sweet}, {Lopez-Sanchez}, {Medling}, {Owers}, {Dubois}, \& {Devriendt}}]{Welker20}
{Welker}, C., {Bland-Hawthorn}, J., {van de Sande}, J., {et~al.} 2020, \bibinfo{title}{{The SAMI Galaxy Survey: first detection of a transition in spin orientation with respect to cosmic filaments in the stellar kinematics of galaxies},} \mnras, 491, 2864, \dodoi{10.1093/mnras/stz2860}

\end{thebibliography}
\bibliographystyle{aasjournalv7}



\end{document}